\documentclass[smallextended]{svjour3}
\journalname{Eur. Phys. J. D}
\usepackage[utf8]{inputenc}
\usepackage[T1]{fontenc}
\usepackage{geometry}
\usepackage{amsmath}
\usepackage{amssymb}
\DeclareUnicodeCharacter{2212}{-}
\DeclareUnicodeCharacter{0301}{\'{e}}
\DeclareUnicodeCharacter{0308}{{\"\i}}
\usepackage[backend=biber,sorting=none,style=phys,eprint=true]{biblatex}
\usepackage{xcolor}
\usepackage{graphicx}
\begin{document}
\title{Attoscience in Phase Space}

\titlerunning{Attoscience in Phase Space}  
\author{H. Chomet         \and
        C. Figueira de Morisson Faria}
\institute{
Department of Physics and Astronomy, University College London
Gower Street, London WC1E 6BT, UK 
}
\date{\today}
\maketitle
\begin{abstract}
 We provide a brief review of how phase space techniques are explored within strong-field and attosecond science. This includes a broad overview of the existing landscape, with focus on strong-field ionisation and rescattering, high-order harmonic generation, stabilisation and free-electron lasers. Furthermore, using our work on the subject, which deals with ionisation dynamics in atoms and diatomic molecules as well as high-order harmonic generation in inhomogeneous fields, we exemplify how such tools can be employed. One may for instance determine qualitatively different phase space dynamics, explore how bifurcations influence ionisation and high-harmonic generation, establish for which regimes classical and quantum correspondence works or fails, and what role different time scales play. Finally, we conclude the review highlighting the importance of the tools available in quantum optics, quantum information and physical chemistry to strong-field laser-matter interaction.


\end{abstract}

\maketitle

\section{Introduction}
\label{sec:intro}

The idea of phase space, in which one may depict all possible states of a dynamical system evolving from \textit{any} initial conditions by trajectories, is extremely powerful. Each phase space trajectory represents the evolution of a system starting from specific initial conditions, with each point corresponding to the state of the system at a specific time. The set of all phase space trajectories thus provide a mapping of all possible ways in which a system may evolve. In particular, the phase space is used for dealing with multidimensional systems, whose description would be much less intuitive otherwise. Examples of such systems are encountered in a wide range of areas, including physics, biology, chemistry and financial models (see \cite{Weinbub2018} for a recent review).  In physics alone, phase space tools are typically used in, for instance, statistical physics, quantum optics, collision theory, particle physics and nonlinear dynamics, and widely employed phase space variables are, for instance, positions and momenta, or angles and angular momenta. Its mathematical origin, dating from 1838, can be attributed to Liouville \cite{Liouville1838}, and its first application to mechanics was made by Jacobi in 1842 \cite{Jacobi1884}. However, the concept of describing the dynamics of a system as a single trajectory moving through multidimensional space was developed many decades later by Poincar\'e \cite{Poincare1993} (for a historic review on the subject see \cite{Nolte2010}). 

The quantum phase space was introduced much later, by Wigner, together with the quasiprobability distribution named after him \cite{Wigner1932}. Since then, quantum phase space distribution functions, constructed using non-commuting operators, have become widespread. A key advantage is that they allow one to employ complex-number functions instead of dealing with operators. Furthermore, they provide valuable insight in quantum-classical correspondence \cite{Heller1976}, within the constraints posed by the uncertainty principle and its generalisations.  However, there are different phase space distribution functions, whose applicability may suit specific problems better than others. This ambiguity stems from the fact that there are different rules for associating non-commuting operators to scalar variables \cite{Hillery1984,Lee1995}; for pioneering work exploring operator ordering in connection with quasiprobability distributions see also \cite{Cahill1969,Agarwal1970}. For instance, due to their smooth behavior, Husimi distribution functions are popular in the context of nonlinear systems and classical chaos \cite{Takahashi1985}, while Wigner quasiprobability distributions, due to the information they provide on non-classical effects and quantum corrections, are widely used in quantum optics \cite{schleich2011quantum,Barnett2005}.  Other applications of the Wigner function include  optical propagation in waveguides \cite{Alonso2011}, and the computation of angular momentum states \cite{Agarwal1981}, which can also be used to model two-level atoms \cite{Dowling1994,Czirjak1996,Benedict1999}. The Glauber-Shudarshan $\mathcal{P}$ functions \cite{Glauber1963a,Sudarshan1963} are also hugely popular in quantum optics, as they are very convenient for normal-ordered products of creation and annihilation operators.

Quantum phase space distribution functions play a major role in quantum optics \cite{schleich2011quantum,Barnett2005} and quantum information \cite{Braunstein2005,Serafini2017}. This popularity has been triggered by the description of the electromagnetic field modes as quantum harmonic oscillators, for which distribution functions have been especially tailored (see, e.g., the discussion in  \cite{Hillery1984,Lee1995,Glauber1963a,Sudarshan1963}), and the central interest in the definition of non-classical states of light \cite{Dodonov2002}. Furthermore, due to being formulated in terms of density matrices, quasiprobability densities are well suited for investigating decoherence and the influence of the environment \cite{Garraway1994,Garraway1994a,Benedict1999}. Quasiprobability distributions have also been explored in connection with logical gates \cite{Veitch2012,Ketterer2016} and their classical simulation \cite{Raussendorf2020}, and coherent-state superposition \cite{Vogel1989,Garraway1993}.

Other traditional areas in which the quantum phase space is widely used are those dealing with large systems \cite{Polkovnikov2010}, such as chemical physics \cite{Miller2001,Miller2005} and cold gases \cite{Blakie2008}. In this case, the huge amount of degrees of freedom makes a full quantum-mechanical treatment prohibitive. Therefore, crucial questions are what degrees of freedom need full treatment and which ones can be approximated, what kind of fluctuations and deviations from the classical picture are expected, and whether there are semiclassical limits one can take into consideration without losing essential information about the system's dynamics.

In the study of complex molecular systems, for instance, it is common to apply mixed classical-quantum methods, which describe less relevant degrees of freedom classically, more relevant degrees of freedom quantum mechanically, and couple them via effective potentials \cite{Goran1988,Wang1998}. One may also consider systems coupled to baths, whose dynamics are simplified \cite{Caldeira1983}. Alternatively, one may develop semiclassical methods, in which swarms of classical trajectories are employed to construct quantum propagators (see, e.g.,  \cite{Heller1975,Miller1975,Heller1981,Herman1984,Shalashilin2004} and the reviews \cite{Miller2001,Werther2021}). Further approximations may then be applied, such as the smoothening of highly oscillatory terms \cite{Heller1976}, and linearised semiclassical approximations, in which the main contributions stem from trajectories whose phase space coordinates are close enough \cite{Miller1975}. Truncated Wigner approximations are also widely used to describe chemical reactions (for an early example see \cite{Miller1975}).

Furthermore, there are many perturbative approximations which incorporate quantum fluctuations around classical limits, such as the semiclassical or truncated Wigner approximation (TWA). For an early discussion of phase space methods in which stochastic evolution equations, including the TWA, are applied to Bose-Einstein Condensates see \cite{Steel1998}.  The key idea is to embed quantum fluctuations in the initial quasiprobability distributions, which are then evolved classically. This means that quantum corrections appear only through initial conditions. Classical evolution is desirable for large systems as, typically, classical equations of motion scale linearly with its degrees of freedom, while quantum methods scale exponentially. The TWA has been explored in the context of Bose condensed gases perturbed from thermal equilibrium \cite{Sinatra2002} in a wide range of scenarios such as the evolution of macroscopic entangled states \cite{Polkovnikov2003}, quantum fluctuations in condensate oscillations \cite{Polkovnikov2004}, non-classical effects arising from the splitting of condensates \cite{Isella2005,Isella2006} and vortices in reflecting Bose-Einstein condensates \cite{Scott2006}. It works for weakly perturbed systems. 

In contrast, in strong-field laser-matter interaction and attosecond science, 
the phase space picture and its tools have not become mainstream.
This is actually surprising for the following reasons:
\begin{itemize}
    \item \textit{Classical and semi-classical trajectory-based methods have been used as interpretational tools for quantum effects since over two decades, and helped establish the key paradigms in the field. }Therein, laser-induced rescattering and recombination of an electron with its parent ion play a vital role by providing an intuitive interpretation of many strong-field phenomena (see, e.g., \cite{Corkum1993,Lewenstein1994,Salieres2001} and the special issue \cite{Schafer2017}). If recombination with a bound state occurs, the kinetic energy acquired by the electron in the continuum is released in form of high-order harmonic radiation, while rescattering will lead to high-energy photoelectrons. If the returning electron rescatters elastically, high-order above-threshold ionisation (ATI) will occur \cite{Milos2006,Becker2018}. Alternatively, if,  upon recollision, it passes on part of its kinetic energy to the core, it may release other electrons, leading to nonsequential double and multiple ionisation \cite{Faria2011,Becker2012}.  This orbit-based picture has been hugely successful in describing the physics of the problem, with far reaching consequences. For instance, the shapes of the high-harmonic and high-order ATI spectra, with a long plateau followed by sharp cutoffs whose energy positions are a multiple of the ponderomotive energy $U_p$ \cite{Lewenstein1994,Becker2018}, proportional to the driving-field intensity, can be explained using the laser-induced rescattering picture. Furthermore, because ionisation and recombination occur at very specific times within a field cycle, they can be used for steering subfemtosecond electron dynamics, for generating attosecond pulses or bursts of electrons (see, e.g. \cite{Paul2001,Hentschel2001}, and the special issue \cite{Kienberger2012}), and for subfemtosecond imaging of matter \cite{Corkum2007,Lein2007,Krausz2009}.  For a given energy, there is usually more than one quantum mechanical pathway along which the electron may interact with the core, so the corresponding transition amplitudes interfere. Quantum interference has many attosecond imaging applications, using high-order harmonic generation or photoelectrons (for reviews see, e.g., \cite{Augstein2012} or \cite{Faria2020}).
    
    \item \textit{The interaction between the continuum and bound states is of vital importance}, as strong-field ionisation and laser-induced rescattering or recombination play a key role in explaining strong-field phenomena \cite{Lewenstein1994,Salieres2001,Milos2006,Faria2011,Becker2018,Faria2020,Amini2019}. Yet, tools that could provide rigorous and/or accurate information about non-classicality or boundaries of bound-continuum dynamics are underused.
\end{itemize}
  Still, some groups have explored phase space dynamics in a strong-field context. This includes stabilisation \cite{Bestle1993,Watson1995,Watson1996,Norman2015}, strong-field ionisation \cite{Czirjak2000,Takemoto2010,Takemo2011,Zagoya2014,Dubois2018,Chomet2019}, high-order harmonic generation \cite{Graefe2012,Kamor2014,Berman2015,Zagoya2016,Berman2018}, laser-induced core dynamics \cite{Medisauskas2015,Busto2018}, rescattering \cite{Geyer2001,Baumann2015,Graefe2012,Czirjak2013,Kamor2014,Wells2020}, nonsequential double ionisation \cite{Lein2001,Eckhardt2001,Sacha2001,Sacha2001b,Sacha2003,Prauzner2005,Prauzner2007,Mauger2009a,Mauger2009,Mauger2010a,Mauger2010,Mauger2010b,Kamor2011,Mauger2012,Mauger2012a,Dubois2020,Dubois2020a}, or in connection with initial-value representations in strong fields \cite{vandeSand1999,vandeSand2000,Zagoya2012,Zagoya2012b,Zagoya2014}. Furthermore, the growing interest in free-electron lasers (FELs) means that the tools employed in quantum optics are being explored in the X-ray and extreme ultraviolet (XUV) regime \cite{Adams2013,Rohlsberger2014}. Examples range from seminal work unifying quantum and classical descriptions of electron dynamics in FELs \cite{Bambini1980,Procida1984} to providing a road map for quantum signatures therein \cite{Carmesin2020}. They include the development of quantum models whose radiation holds the promise of having better quality than that of classical FELS \cite{Bonifacio2005,Piovella2007,Piovella2008,Kling2015,Carmesin2020}.

Overall, the use of phase space has been twofold: either the \textit{classical} phase space was employed to delimit bound-continuum boundaries, highlight regular or chaotic behavior, and analyse different rescattering regimes, or \textit{quantum} phase space distributions have been employed to assess classical or non-classical behavior and provide initial conditions for other methods. Often the intuitive picture obtained by classical methods is compared with the outcome of \textit{ab-initio} computations or other approaches. 

In the present article, we provide a review of attoscience in phase space. We will start with a brief overview of its use in strong-field and attosecond physics (Sec.~\ref{sec:overview}),  with emphasis on the different phenomena and how specific methods, quantum, classical and semiclassical, have been used over the years. Whilst it is not possible to draw a linear timeline, we have grouped such studies according to common physical or methodological aspects, in order to set an overall landscape. Subsequently, in Sec.~\ref{sec:examples}, we delve deeper into our own work, and use it to exemplify how classical and quantum aspects of the phase space may be used in attoscience. We start by providing a brief statement on the methods (Sec.~\ref{sub:methods}) employed and, in the ensuing sections, focus on common strategies rather than phenomenon or publication. This includes using phase space methods to identify different phase space configurations and bifurcations (Sec.~\ref{sec:config}), distinguishing between classical, semiclassical and quantum regimes (Sec.~\ref{sec:clvsquantum}), and performing an in depth analysis of the different time scales that arise in our investigations (Sec.~\ref{sub:timescales}). Finally, in Sec.~\ref{sec:conclusions}, we conclude the article by placing phase space studies into a broader context. In particular, we highlight the importance of bringing the toolkit available in quantum optics to attosecond physics in view of recent trends and developments. 

\section{Overview}
\label{sec:overview}

Historically, the phase space has been applied to  a variety of phenomena in attoscience, along the following research lines: Free-electron lasers and stabilisation, tunneling, rescattering in one-electron systems and correlated multielectron dynamics.  These research lines often overlap, and a key common aspect is to try to understand and control attosecond electron dynamics in greater depth. They are briefly discussed below.

\subsection{Free-electron lasers and stabilisation}
\label{sub:fel}

Phase space tools have been first used in attosecond science and related fields in the 1980s, in the context of free-electron lasers (FELs) \cite{Bambini1980,Procida1984}. These seminal papers aimed at bridging a gap that existed between fully quantum electrodynamic descriptions of electrons in a FEL and widespread classical descriptions of these dynamics. This was an important milestone as the classical description of electrons in a FEL are expected to become inaccurate in the XUV/X ray regime, for which quantum fluctuations are important \cite{Bambini1980,Procida1984}. For that purpose, Wigner quasiprobability distributions were used and it was shown that, in the classical limit, the former descriptions were recovered. Following that, there have been phase space studies to determine the boundary between classical and quantum behaviour \cite{Bonifacio2005,Piovella2007,Piovella2008,Kling2015,Carmesin2020}. These studies have been motivated by the prospects of developing a Quantum FEL, which should exhibit a narrower linewidth and better temporal coherence than its classical counterpart \cite{Bonifacio2009}. In particular, one- \cite{Piovella2007} and three-dimensional \cite{Piovella2008} quantum models based on Wigner functions are presented.  
In \cite{Kling2015}, the phase space was used to establish a quantum regime, in which the system may be approximated by a two-level atom by averaging over fast oscillations. Recently, quantum effects in the FEL electrons and its gain were studied using Wigner quasiprobability distributions and the quantum Liouville equation \cite{Carmesin2020}. 

Further work in the high-frequency regime, in the 1990s, addressed the question of non-classical behavior in atomic stabilisation. Atomic stabilisation stems from the breakdown of Fermi's golden rule for computing ionisation probabilities in strong laser fields. Roughly speaking, stabilisation is the suppression of ionisation with increasing field strength. In the 1990s, it has generated a great deal of controversy, from its definition, to the physical mechanisms behind it and its existence altogether (for reviews see \cite{Burnett1993,Eberly1993,Geltman1995,Gavrila2002}). In this context, the Kramers-Henneberger frame, in which the field time dependence is embedded in the binding  potential, is widely used.  For high driving-laser frequencies, one may define a double-well effective potential, known as the Kramers-Henneberger potential. 
Wigner quasiprobability distributions were employed to assess under what conditions stabilisation was classical or quantum \cite{Bestle1993}. They were compared to classical-trajectory computations and exhibited regions that were attributed to coherent superpositions of a few bound states of the effective Kramers-Henneberger potential, thus highlighting the role of quantum interference \cite{Watson1995}. Further work explored how quantum effects in stabilisation depend on the pulse shape and on the effective Kramers-Henneberger potential by using Wigner and Husimi distributions \cite{Watson1996}. Much later work relates stabilisation to trapped trajectories and  elliptic islands in a chaotic region via a classical phase space analysis, and highlights a hidden short-time nature of stabilisation \cite{Norman2015}. 

\subsection{Tunneling}
\label{sub:tunneling}

In the low-frequency regime, phase space quantum distribution functions have been employed to assess tunneling ionisation dynamics since the early 2000s. Tunneling is crucial for strong-field and attosecond physics, and the question of with what velocity, and at what point in space an electron reaches the continuum, as well as whether one may define finite tunneling times, has attracted a great deal of attention (see, e.g.,  \cite{Eckle2008,Eckle2008Streaking,Pfeiffer2012,Landsman2015,Torlina2015,Ni2018,Sainadh2019,Douguet2018,Hofmann2019,Eicke2020} for a wide range of approaches and points of view). Since the phase space allows for an intuitive view of the classical-quantum correspondence, it is ideally suited to such questions. 

Early studies of tunneling ionisation in phase space observed a tail for the Wigner quasiprobability distribution of a system in a static or quasi-static, low-frequency field.  This tail has been associated with tunnel ionisation as it crosses from a bound phase space region to the continuum through classically forbidden regions \cite{Balazs1990}. Its slope has been employed to define a tunnel trajectory, which was first computed by \cite{Czirjak2000} using an analytical model of a zero-range potential in a static field. Further work, a decade later,  \cite{Graefe2012} investigated how the slope of the Wigner function behaved with regard to the potential being short or long range. Both publications focused on the agreement between the tail of phase space quantum distributions and a classical-trajectory picture, which were shown to match far away from the core. Nonetheless, quantum interference fringes associated with tunneling events at different times were observed in both publications. Close to the core, the tail of the Wigner function follows the separatrix and crosses into the continuum either via over the barrier or tunnel ionisation \cite{Zagoya2014}. Recently, Wigner quasiprobability distributions have been employed to reconstruct the tunnel exit, which is an important parameter in determining the tunneling time \cite{Hack2019}. Further work by the same group addresses the influence of quantum interference and over-the-barrier ionisation on classical-quantum correspondence when an electron is freed into the continuum \cite{Hack2021}. 

For systems with more than one centre, such as in diatomic molecules, Wigner quasiprobability distributions have been employed in the context of enhanced ionisation \cite{Takemoto2010,Takemo2011,Chomet2019}. Roughly speaking, enhanced ionisation means that, for specific internuclear separations, ionisation rates in a stretched molecule are considerably higher than those in an atom with a similar ionisation potential \cite{Zuo1995}.  
In \cite{Takemoto2010,Takemo2011} it has been shown that there are intra-molecular momentum gates in phase space, which facilitate population transfer within the molecule and to the continuum. They have been attributed to the non-adiabatic response of the molecule to a low frequency field. Further work, however, showed that the momentum gates are intrinsic to the molecular system, and exist even for static fields, or no fields at all. Thereby, quantum interference plays an important role by providing a bridge for quasiprobability to flow from one molecular well to the other \cite{Chomet2019}, and the frequencies can be estimated for double-well potential models \cite{Kufel2020}. Further studies of non-adiabatic effects and bifurcation in strong-field ionisation were conducted in \cite{Dubois2018}.

Tunneling dynamics in strong fields has also been looked at in the context of initial-value representations (IVRs) \cite{Spanner2003,Zagoya2014}. In initial-value representations, the boundary problems that arise in semiclassical theory are replaced by averages over initial phase space coordinates, which are used to construct wave packets. These wave packets are then evolved in time, guided by ensembles of classical trajectories.  
 IVRs are employed in many areas of science, for instance quantum chemistry, chaos and nonlinear dynamical systems, and are very powerful approaches due to their scalability and absence of cusps and singularities. For key references see, e.g.,  \cite{Heller1975,Miller1975,Heller1981,Herman1984,Shalashilin2004} and the reviews \cite{Miller2001,Werther2021}.  However, there has been considerable debate whether these approaches can be used to model tunnel ionisation, as they employ ensembles of real classical trajectories to construct wave packets \cite{Keshavamurthy1994,Grossmann1995,Kay1997,Maitra1997}.Tunneling may manifest itself in position space, as transmission, or in momentum space, as above-the-barrier reflection, and it not being accounted correctly will cause semiclassical IVRs to degrade for longer times  \cite{Maitra1997}. Nonetheless, because the top of a potential barrier can be approximated by an inverted harmonic oscillator, tunneling has been found to work well near this threshold \cite{Balazs1990}. To deal with this, one may either focus on rescattering only and place the initial electronic wave packet away from the core \cite{vandeSand1999,vandeSand2000}, or employ short times and IVRs with effective potentials that account for quantum corrections, such as the Coupled Coherent State (CCS) method \cite{Zagoya2014}.  
The phase space has also been employed to develop path integral approaches \cite{Milosevic2013JMP} that incorporate the residual potentials and the driving field on equal footing, such as the Coulomb Quantum Orbit Strong-Field Approximation (CQSFA) \cite{Lai2015,Maxwell2017,Maxwell2018,Maxwell2018b} and the semiclassical two-step
(SCTS) model \cite{Shvetsov-Shilovski2016,Shvetsov-Shilovski2019}, with emphasis on quantum interference and photoelectron holography. For a review see \cite{Faria2020}.

\subsection{Rescattering in one-electron systems}
\label{sub:resc}

Because most strong-field phenomena can be explained as laser-induced rescattering, one must understand how it manifests itself in phase space. Although structures associated with rescattering have already been identified in \cite{Czirjak2000}, closer scrutiny happened only in the 2010s. In \cite{Kull2012}, distinct interference patterns in Wigner quasiprobability distributions have been associated with different types of intra-cycle electron scattering and above-threshold ionisation. This has been extended in \cite{Baumann2015} in order to assess lower impact velocities, and to compute the bound-state population using phase space criteria.  Therein, phase space signatures of channel closings have also been identified. Further work has investigated the connection between rescattering and entanglement \cite{Czirjak2013}.

Rescattering in phase space has also been studied in relation to other phenomena. For instance, in \cite{Graefe2012}, a phase space analysis of rescattering in conjunction with high-order harmonic generation (HHG) was performed employing Wigner and Husimi distribution functions. It was shown that the rescattering wave packet exhibits a chirp, which can be extracted from the Wigner quasiprobability distribution at the position of rescattering. The HHG temporal profile given by the Wigner function strongly resembles that obtained by other means such as windowed Fourier transforms. 
Recent work has focused on a systematic analysis of the orbit-based rescattering picture for tunneling, rescattering and HHG using Wigner functions with spatial windows in reduced-dimensionality models, and effective Wigner functions for multidimensional systems in order to facilitate the interpretation of more intricate dynamics \cite{Wells2020}.

Different types of orbits and their role in HHG \cite{vandeSand1999} and ATI \cite{vandeSand2000} have also been investigated using initial-value representations. It was found that irregular orbits play an important role in forming the HHG plateau.  Furthermore, phase space tools have been employed to identify regions of dominant, integrable Hamiltonians, which led to HHG spectra with excellent agreement with ab-initio methods \cite{Zagoya2012,Zagoya2012b}. A key challenge in modeling HHG is that it is a coherent process that relies on the periodicity of the field. This implies that any dephasing associated with the degradation of the time evolution determined by the IVR will affect the harmonic spectra. This will play a key role if the wave packet is initially bound as tunnel ionisation will be important in this case (for discussions see \cite{Zagoya2014,Symonds2015}).

Subsequently, a purely classical perspective into how the presence of the Coulomb potential affects recollisions in strong fields, as related to the high-harmonic spectra, is provided in a series of publications \cite{Kamor2014,Berman2015,Berman2018}. Therein, classical phase space arguments have been used to show that Coulomb focusing enhances delayed recollisions and increase their energy. These recollisions occur along periodic orbits whose energy are higher than the well-known value of $3.17U_p$ \cite{Berman2015}. Nonetheless, in \cite{Kamor2014}, a fully classical method that considers the Coulomb potential in the continuum is employed to explain why the standard cutoff law works. A set of periodic orbits stemming from a resonance with the field are linked to laser-induced recollision, whose maximal energy approaches the standard HHG cutoff in the high-intensity limit. Good agreement with the \textit{ab-initio} solution of the time-dependent Schr\"odinger equation is observed. Further work explores the extension of the cutoff upon macroscopic propagation \cite{Berman2018}. The phase space has also been employed by us in \cite{Zagoya2016} to extract different instantaneous configurations and time scales for HHG in inhomogeneous fields. 

\subsection{Correlated multielectron processes}
\label{sub:nsdi}

In addition to one-electron problems, since the early 2000s, the phase space has been used to explore non-trivial features in correlated multielectron processes. This extends from laser-induced nonsequential double ionisation \cite{Faria2011}, which is the archetypical example of electron-electron correlation in intense laser fields \cite{Lein2001}, to the temporal profile of autoionisation dynamics in Helium \cite{Busto2018}. For instance, in \cite{Lein2001} Wigner quasi-probability distributions associated with the centre-of mass coordinates of a two-particle system have been compared to the outcome of a mean-field theory in order to identify signatures of rescattering and electron-electron interaction. NSDI has also been modeled for the Helium atom using IVRs beyond reduced-dimensionality models \cite{Shalashilin2008,Kirrander2011}. In particular an alternative version of the Coupled Coherent States (CCS) method that incorporates the exchange symmetry of fermionic particles, the fermionic CCS, has been successfully applied in this context \cite{Kirrander2011}.   

A whole line of research has been devoted to investigating NSDI in a fully classical framework. In NSDI, an outer electron reaches the continuum, is brought back by the field and transfers part of its kinetic energy to an inner electron. These recollision dynamics are quite rich and can be interpreted using tools from phase space, the theory of non-linear dynamical systems,  and effective Hamiltonians for each of the electrons involved. These reduced, integrable Hamiltonians were first defined in \cite{Mauger2009a,Mauger2009} for NDSI. Subsequently, the role of multiple recollisions on the efficient energy transfer in NSDI has been investigated using symplectic maps and similar approaches to those used in kicked Rydberg atoms, and a road map has been provided for identifying different NSDI mechanisms in \cite{Mauger2010a,Mauger2010}. Further work by the same group has focused on an in-depth analysis of recollision excitation with subsequent ionisation (RESI) in terms of resonances and their proximity to periodic orbits \cite{Mauger2012,Mauger2012a}. In particular, a sticky region in phase space arises due to the interplay of the external field and the binding potential. This region traps trajectories before ionisation, leading to time delays for the second electron. A detailed analysis of the types of periodic orbits, resonance conditions and distinct sources of chaos is provided in \cite{Mauger2012a}. Interestingly, if reduced-dimensionality models are used, oscillations in the RESI yield as functions of the laser intensity are reported and attributed to resonances. However, these oscillations are washed out if more degrees of freedom are incorporated, due to chaotic transverse dynamics and additional resonances. These oscillations are distinct from those attributed to quantum interference in RESI \cite{Hao2014,Maxwell2015,Maxwell2016,Quan2017}. Further work is related to NSDI in  bichromatic, linearly polarised fields \cite{Kamor2011}, and the dynamics of recollisions in fields with circular polarisation \cite{Mauger2010b,Dubois2020,Dubois2020a}. In \cite{Mauger2010b} it is shown that, in contrast to previous assumptions, recollisions may occur in circularly polarised fields, by analysing the system's dynamics in a rotating frame. The physical mechanism is similar to that leading to ionisation in Rydberg atoms in microwave fields. A decade later, this topic is revisited and several associated time scales  are analysed in detail \cite{Dubois2020,Dubois2020a}. In particular, a recollision mechanism taking place over many field cycles is reported. 

One may also employ classical models and dynamical systems tools to determine modified threshold laws for correlated multielectron ionisation that account for the presence of an external field \cite{Eckhardt2001}. This approach starts by identifying similarities with its field-free counterpart \cite{Wannier1953},  considering the two electrons in an excited compound and identifying the relevant subspaces for which correlated double and multiple ionisation may occur. This information can then be used to construct effective reduced Hamiltonians for the subspace of interest, identify existing symmetries and possible electron escape configurations, and determine which of the latter will prevail. This has been done for nonsequential double \cite{Sacha2001,Prauzner2005}, triple \cite{Sacha2001b} and multiple \cite{Sacha2003} ionisation. One can also use this method as guidance for defining effective reduced-dimensionality quantum models, so that the actual dynamics are preserved as faithfully as possible, without introducing artificial constraints or correlations \cite{Prauzner2007}.

Another manifestation of electron-electron correlation is autoionisation. Thereby, the quantum interference between a direct and a quasi-resonant pathway is of extreme importance. Wigner quasi-probability distributions in the time-energy domain are used to study this interference and disentangle these pathways in a transient process \cite{Busto2018}. They provide an advantage over other methods used in time-frequency analysis, for exposing non-classical behavior in a much more explicit way. A further issue is that, in classical-trajectory models, autoionisation  manifests itself as an artifact that renders the system unstable. These problems have been overcome in \cite{Geyer2001,Geyer2003}, which employ quasiprobability densities and phase space arguments to develop classical-trajectory models that do not exhibit these shortcomings and are consistent with their quantum-mechanical counterparts. 

\section{Selected examples}
\label{sec:examples}

In the following, we provide a few selected examples from our own work of how phase space tools can be used to model strong-field dynamics and extract information that may not be available by other means.  These are used to classify regions with qualitatively different behaviors, and understand how processes such tunneling and rescattering unfold. For details, we refer to the original publications \cite{Zagoya2014,Zagoya2016,Chomet2019,Kufel2020}. We also intend to go beyond these articles, by bringing together aspects that have not been emphasised so far. 

\subsection{Methods}
\label{sub:methods}

We employ both classical and quantum-mechanical tools. We focus on simplified,  reduced-dimensionality models in which a single spatial dimension is taken into consideration. They provide a transparent, yet accurate picture of the system's dynamics for linearly polarised fields. We also use atomic units. One should note, however, that the choice of lower-dimensional models is not trivial in attosecond physics, as one must ensure that no artifacts such as spurious symmetries and correlations are introduced, with regard to realistic, multidimensional models. For correlated electron dynamics, this is briefly mentioned in Sec.~\ref{sub:nsdi}. For one-electron problems, a discussion of how to keep the one-dimensional dynamics and observables as close as possible to a three-dimensional system is given in  \cite{Majorosi2018,Majorosi2020}.

\subsubsection{Classical phase space dynamics}
\label{sec:classicalphspc}

Classically, the phase space dynamics are described by Hamilton's equations
\begin{equation}
\dot{x} = p = \frac{\partial H_{\mathrm{cl}}(p,x)}{\partial p} 
\label{eq:phaseSpaceX}
\end{equation}
\begin{equation}
\dot{p} = - \frac{\partial V_{\mathrm{eff}}}{\partial x} = -\frac{\partial H_{\mathrm{cl}}(p,x)}{\partial x},
\label{eq:phaseSpaceP}
\end{equation}
where $x$ and $p$ are the position and canonically conjugate momentum, respectively, and the classical Hamiltonian $H$ is defined by 
\begin{equation}
    H_{\mathrm{cl}}(p,x)=\frac{p^2}{2}+V_{\mathrm{eff}}(x).\label{eq:Hclassical}
\end{equation}
In Eq.~(\ref{eq:Hclassical}), $V_{\mathrm{eff}}=V(x)+V_l$ is the effective potential determined by the external electric field acting on the electronic wave packet. The physical picture of a time-dependent effective potential corresponds to the length gauge and the dipole approximation, which are used in this work.  Thereby, $V_l$ is the potential energy determined by the laser field, which for a field without spatial dependence, reads $V_l=x\mathcal{E}(t)$.

Here we take the binding potential $V(x)$ to be either soft-core
\begin{equation}
V_{\mathrm{sc}}(x)=-\frac{1}{\sqrt{x^2+\lambda}},  \label{eq:softcore}
\end{equation}
with $\lambda=1$ constant, or as the short-range Gaussian
potential
\begin{equation}
V_{G}(x)=-\exp(-\lambda x^2), \label{eq:gaussian}
\end{equation}
where $\lambda=1/2$. For diatomic molecules we consider 
\begin{equation}
    V(x)=V_{\mathrm{sc}}(x-R/2)+V_{\mathrm{sc}}(x+R/2).
    \label{eq:molpot}
\end{equation}
The field is assumed to be either static, i.e., $\mathcal{E}(t)=\mathcal{E}_0$, or as a linearly polarised monochromatic wave of frequency \(\omega\) and phase \(\phi\)
\begin{equation}
    \mathcal{E}(t)=\mathcal{E}_0 \cos(\omega t + \phi).
    \label{eq:efield}
\end{equation}
Specifically in the work associated with inhomogeneous fields we introduce the spatial dependence of the laser field by considering

\begin{equation}
  \mathcal{E}(x,t) = (1+\beta x)\mathcal{E}(t),
   \label{eq:inhom}
\end{equation}
and the associated inhomogeneous effective potential
\begin{equation}
    \tilde{V}_{\mathrm{eff}}=V(x) + (x+0.5\beta x^2) \mathcal{E}(t),
    \label{eq:inhomV}
\end{equation}
where $\beta$ is an inhomogeneity parameter\footnote{The present expression has been corrected by a factor two with regard to that in \cite{Zagoya2016}. However, we have verified that this does not change the main conclusions in the original publication.}. This approximation has been widely employed in the literature \cite{ciappina2012,Ciappina2012b,shaaran2012,Yavuz2012,shaaran2013,Hekmatara2014,Luo2014} contingent on a small inhomogeneity parameter. 

This classical approach is used to determine bound and continuum phase space regions, identify fixed points and analyse ensembles of trajectories under different conditions. This is done by employing key concepts of the theory of dynamical systems, some of which  are briefly stated here for the sake of self consistency. For a more detailed and rigorous discussion please see e.g., \cite{Arrowsmith1992}. Solutions of Eqs. (\ref{eq:phaseSpaceX}) and (\ref{eq:phaseSpaceP}) which stay the same for all times, that is, for which $\dot{x}=\dot{p}=0$, are fixed points. For conservative Hamiltonian systems of the form (\ref{eq:Hclassical}), such as a model atom in a static field ($\mathcal{E}(t)=\mathcal{E}_0$), one may show that fixed points are centres or saddles. In phase space, these fixed points are located at $ (x_f,p_f)=(x_s,0)$, where $x_s$ is the value of the coordinate $x$ for which the effective potential $V_{\mathrm{eff}}$ or $\tilde{V}_{\mathrm{eff}}$ is stationary. Centres and saddles are given by the minima and maxima of the effective potential, respectively. Centres are attractive and surrounded by closed orbits, while saddles are semi-stable and help delimit qualitatively different dynamical regions in phase space. For that reason, phase space trajectories crossing saddles are called \textit{separatrices}. For a time-dependent field, this line of argument is approximate, but it can be used if its frequency is low enough for the quasi-static picture to hold. This is the case in the present work.

\subsubsection{Time-dependent Schr\"odinger equation and initial wave packet}

As a benchmark, we use the full solution of the time-dependent Schr\"odinger equation (TDSE),
\begin{equation}
\mathrm{i}\partial_t |\Psi(t)\rangle=\hat{H}|\Psi(t)\rangle, \label{eq:tdsegeneral}
\end{equation}
where the length-gauge Hamiltonian reads
\begin{equation}
\hat{H}=\frac{\hat{p}^2}{2}+\hat{V}_{\mathrm{eff}(\hat{x})},\label{eq:Hoperator}
\end{equation}
with $V_{\mathrm{eff}}(x,t)$ being defined as above and the hats denoting operators. We solve the TSDE in the position space,
\begin{equation}
    i \partial_t\Psi(x,t)=\left(-\frac{1}{2}\frac{\partial^2}{\partial x^2}+V_{\mathrm{eff}}(x,t)\right)\Psi(x,t),
    \label{eq:tdse}
\end{equation}
where $\Psi(x,t)$ is the time-dependent wave function. Depending on the problem at hand, we choose different initial wavefunctions $\Psi(x,0)$ and potentials $V_{\mathrm{eff}}(x)$. 

We will approximate the initial wave function by  Gaussian wave packets 
\begin{equation}
\Psi(x,0)=\langle x|\Psi (0) \rangle = \left( \frac{\gamma}{\pi} \right)^{\frac{1}{4}}\exp \left\{-\frac{\gamma}{2}(x-q_{0})^{2}+ip_{0}(x-q_{0})  \right\}
\label{eq:Psi0}
\end{equation}
of width $\gamma $ centred at vanishing initial momentum $p_0=0$ and initial coordinate $q_0$, or coherent superpositions thereof.
Specifically for diatomic molecules, we will consider $\gamma = 0.5$ a.u. and $q_0=-R/2$ or $q_0=R/2$, in which cases the initial wave functions are given by $\Psi_{\mathrm{down}}(x,0)$ or $\Psi_{\mathrm{up}}(x,0)$, respectively.

The delocalised wave function is taken to be the symmetric coherent superposition 
\begin{equation}
\Psi_{\mathrm{cat}}(x,0) = \frac{\Psi_{\mathrm{down}} (x,0)+\Psi_{\mathrm{up}}(x,0)}{\sqrt{\int \left[\Psi_{\mathrm{down}} (x,0) + \Psi_{\mathrm{up}}(x,0)\right]^2dx}}.
\label{eq:statcat}
\end{equation}

The time-dependent wave function is employed to compute quantum distribution functions and observables discussed in the next section. It will also be used to calculate the time dependent autocorrelation function
\begin{equation}
a(t) = \int \Psi^* (x,t) \Psi (x,0) dx,
\label{eq:autocorrelation}
\end{equation}
as well as the ionisation rate
\begin{equation}
\Gamma=-\ln\left(\frac{|\mathcal{P}(T_f)|^{2}}{|\mathcal{P}(0)|^{2}}\right)\frac{1}{T_f},
\label{eq:ionisation}
\end{equation}
 from an initial time  \(t=0\) to a final time \(t=T_f\), where
\begin{equation}
\mathcal{P}(t)=\int^{+\infty} _{-\infty}\Psi^*(x,t)\Psi(x,t)dx.
\label{eq:normWF}
\end{equation} 
This definition of ionisation rate was used in the seminal paper \cite{Zuo1995} in the context of enhanced ionisation of molecules, and in our previous publications \cite{Chomet2019,Kufel2020}. 
\subsubsection{Quantum distribution functions}

The time-dependent wave function is used as input in the Wigner quasiprobability distribution
\begin{equation}
W(x,p,t)= \frac{1}{\pi} \int_{\infty}^{-\infty}d\xi\Psi^{*}(x+\xi,t)\Psi(x-\xi,t)e^{2ip\xi}. 
\end{equation}
This function is always real. However, it exhibits both positive and negative values, hence the name ``quasiprobability''. This, among other features, makes its interpretation as a simple probability distribution difficult. This is why it is often paired with a study of classical trajectories in phase space. 
On the other hand, its deviations from probability densities  can be used to define non-classicality. For instance, a widespread definition used in quantum optics is to seek regions for which $W(x,p,t)$ is negative and classify them as non-classical \cite{Kenfack2004}. A more restrictive definition is based on the quantum Liouville equation \cite{schleich2011quantum}.
\begin{equation}
\left( \frac{\partial}{\partial t} + \frac{p}{M} \frac{\partial}{\partial x}-\frac{dV_{\mathrm{eff}}(x)}{dx} \frac{\partial}{\partial p}\right) W(x,p,t)= Q(x,p,t),
\label{eq:liouville}
\end{equation}
where
\begin{equation}
Q(x,p,t)= \sum_{l=1} ^{\infty} \frac{(-1)^{l}(\hbar/2)^{2l}}{(2l + 1)!} \frac{d^{2l+1}V_{\mathrm{eff}}(x)}{dx^{2l+1}} \frac{\partial^{2l+1}}{\partial p^{2l+1}} W(x,p,t)
\label{eq:quantumcorr}
\end{equation}
are the quantum corrections to the classical Liouville equation. The quantum Liouville equation (also known as Moyal equation) may also be written more compactly as
\begin{equation}
    \frac{\partial W(x,p,t)}{\partial t}= -\left\{\{  W(x,p,t), H(x,p,t)\right \}\},
    \label{eq:LiouvMoyal}
\end{equation}
where $H(x,p,t)$ is the system's Hamiltonian and $\{\{ \cdot \}\}$ give a Moyal bracket \cite{Moyal1949,Groenewold1946}\footnote{Moyal brackets map non-commuting operators to functions in phase space and have been used in a wide range of problems; for instance, one of us applied them to non-Hermitian Hamiltonian systems \cite{Faria2006}.}. In the classical limit (setting \(\hbar = 0\)), Eqs.~(\ref{eq:liouville}) and (\ref{eq:LiouvMoyal}) become the classical Liouville equation, so that \(Q(x,p,t) = 0\) and the right-hand side of Eq.~(\ref{eq:LiouvMoyal}) will be given by a Poisson bracket. In this case, the Wigner quasiprobability distribution will evolve like a classical entity. This is a useful tool for distinguishing between regimes in which quantum interference is present, but evolves classically by, for instance, following classical separatrices, and those truly quantum regimes with no classical counterpart. A widespread approach in quantum optics and cold gases, known as the truncated Wigner approximation (TWA), is to consider the classical Liouville equation with stochastic quantum corrections. This is many times required in order to deal with large systems. The TWA was first used in the context of Bose-Einstein condensates in \cite{Steel1998}; for reviews see \cite{Blakie2008,Polkovnikov2010}, and is closely related to the linearised semiclassical IVR \cite{Miller1975,Miller2001}.  Nonetheless, in some instances it may be nontrivial to compute a classical limit for the quantum Liouville (Moyal) equation (see \cite{Heller1976} for an early discussion). 

One should note that quantum distribution functions may be defined using any two variables corresponding to incompatible observables, such as time and frequency. For instance, Wigner-type time-frequency distributions were employed to study HHG \cite{Kim2001}, different regimes in ATI \cite{Guo2012} and  autoionisation  \cite{Busto2018}. This approach bears some similarity with the use of windowed Fourier transforms, which is much more widespread and has been used since the 1990s to infer time profiles of harmonics and extract electron return times from ab-initio computations (see, e.g., Refs.~\cite{Antoine1995,Faria1997,Faria1998,deBohan1998,Faria1999,Tong2000} for early studies or  Refs.~\cite{Chirila2010,ciappina2012,Ciappina2012b,Wu2013a,Wu2013b} for more recent publications).
\subsubsection{Gabor and Fourier transforms}

In our work we use the Gabor transform to compute time-resolved HHG. The time-resolved spectra are given by $\chi_G(\Omega,t)=|d_G(\Omega,t)|^2$, where $\Omega$ is the harmonic frequency,
\begin{equation}
d_{\mathrm{G}}(\Omega,t)=\int\mathrm{d}t'\,d(t')\mathrm{e}^{-\mathrm{i}\Omega t'-(t'-t)^2/2\sigma^2},\, 
\label{eq:Gabor}
\end{equation}
with $\sigma=1/3\omega$, is a windowed Fourier transform with a Gaussian window function, and
\begin{eqnarray}
d(t)=&-&\langle\Psi(t)|\frac{\partial V_{\mathrm{eff}}(x,t)}{\partial x}|\Psi(t)\rangle \notag \\
=&-&\int \Psi^*(x,t)\frac{\partial V_{\mathrm{eff}}(x,t)}{\partial x}\Psi(x,t)\mathrm{d}x,
\label{eq:acceleration}
\end{eqnarray}
is the dipole acceleration \cite{Sundaram1990,Burnett1992,Krause1992}.
The standard HHG spectrum, for which all temporal information is lost, is computed as $\chi(\Omega)=|d(\Omega)|^2$, with
\begin{equation}
d(\Omega)=\int\mathrm{d}t\,d(t)\mathrm{e}^{-\mathrm{i}\Omega t},
\end{equation}
and is recovered by setting $\sigma \rightarrow \infty$ in Eq.~(\ref{eq:Gabor}).

\subsubsection{Initial value representations (IVRs)}
 \label{subsub:IVRs}

We have also employed initial-value representations (IVRs), where the time-dependent wave function is constructed using a basis of Gaussian wave packets in phase space guided by a trajectory-based grid. IVRs exhibit many advantages, such as providing an intuitive picture in terms of electron orbits, accounting for external fields, binding potentials and quantum interference. Furthermore, they are applicable to large systems due to the numerical effort involved not scaling exponentially with the number of degrees of freedom. Specifically, we employ the Herman Kluk (HK) propagator \cite{Herman1984} and the Coupled Coherent State (CCS) representation \cite{Shalashilin2004}.  Coherent states are widely used in chemical physics and quantum optics for behaving in a quasi-classical way. This allows descriptions in terms of functions of complex numbers, and using the phase space approaches mentioned in this work. The most common coherent states are those of the quantum harmonic oscillator. They have been first introduced by Schr\"odinger in \cite{Schrodinger1926}, and systematically studied by Glauber  \cite{Glauber1963a} and Sudarshan \cite{Sudarshan1963} in the context of quantum optics. Coherent states have also been widely used to construct IVRs guided by Gaussians in phase space, such as the HK propagator \cite{Herman1984}, the CCS representation \cite{Shalashilin2004}, the linearised semiclassical IVRs \cite{Miller1975} and the Frozen Gaussian Approximation \cite{Heller1981}. For a detailed discussion and more coherent-state based IVRS see the recent review  \cite{Werther2021}.\footnote{One should note that there are other types of coherent states, such as the  Barut–Girardello coherent states \cite{Barut1971}, or Gazeau–Klauder \cite{Klauder1960,Klauder1963,Klauder1963a,Gazeau1999} coherent states. For reviews on different types of coherent states see, e.g.,  \cite{Zhang1990,Dodonov2002} and \cite{Dey2018}.}

For the Herman Kluk propagator, we consider a state vector
\begin{equation}
|\Psi_{\mathrm{HK}}(t)\rangle=\int\hspace{-0.2cm}\int%
|q,p\rangle R(t,q_0,p_0)\langle q_0,p_0|\Psi(0)\rangle\mathrm{e}^{\mathrm{i}
S_{\mathrm{cl}}}\frac{\mathrm{d} q_0\mathrm{d} p_0}{2\pi}, \label{PsiHK}
\end{equation}
where $|q,p\rangle$ represents a coherent state in phase space and $q_0,p_0$ the initial phase space coordinates. This state corresponds to a Gaussian wave packet, see Eq.~(\ref{eq:Psi0}). 

The prefactor
\begin{equation}
R(t,q_0,p_0)=\frac{1}{2^{1/2}}\left(m_{pp}+m_{qq}-\mathrm{i}\gamma m_{qp}+\frac{
\mathrm{i}}{\gamma}m_{pq}\right)^{1/2}\label{eq:prefactor_HK}
\end{equation}
is given in terms of the elements $m_{uv}=\partial u/\partial v_0$ of the monodromy matrix, which is composed of the derivatives of the final phase space variables with regard to their initial values, and 
\begin{equation}
S_{\mathrm{cl}}(q,p)=\int\left(p\dot{q}-H_{\mathrm{cl}}(p,q)\right)dt, \label{eq:action_HK}
\end{equation}
is the semiclassical action,
where $H_{\mathrm{cl}}(p,q)$ is the classical Hamiltonian (see Eq.~(\ref{eq:Hclassical}) in this work). However, for clarity,  when using IVRs we employ a slightly different notation than in the rest of the paper. This is done in order to distinguish between phase space Gaussian coherent states and their position-space representations (see Eq.~(\ref{eq:ansatz}) below).

For a Gaussian initial wave packet,
\begin{equation}
\hspace*{-0.8cm}\langle q_0,p_0|\Psi(0)\rangle=\exp\bigg\{-\frac{\gamma}{4}(q-q_0)^2-\frac{1}{4\gamma}(p-p_0)^2+\frac{\mathrm{i}}{2}(p+p_0)(q_0-q)\bigg\}.
\label{eq:initialphasespace}
\end{equation}

For the CCS method, we write the time-dependent wave function as a superposition of time-dependent, nonorthogonal Gaussian coherent states (CS) $|z\rangle=|z(t)\rangle$, defined as 
\begin{equation}\hat{a}|z\rangle=z|z\rangle \quad \mathrm{and} \quad \langle z|\hat{a}^{\dagger}=\langle z|z^*,\end{equation}
where  $\hat{a}^{\dagger},\hat{a}$ are the creation and annihilation operators, respectively, 
and whose eigenvalues are parametrised as functions of the phase space coordinate as
\begin{eqnarray}
z&=&\sqrt{\frac{\gamma}{2}}q+\frac{\mathrm{i}}{\sqrt{2\gamma}}p, \notag \\
z^{*}&=&\sqrt{\frac{\gamma}{2}}q-\frac{\mathrm{i}}{\sqrt{2\gamma}}p.
\label{eq:CS}
\end{eqnarray}
The time-dependent state reads
 \begin{equation}|\Psi (t)\rangle=\int |z\rangle D_{z}(t)\mathrm{e}^{\mathrm{i}S_z}\frac{\mathrm{d}^2z}{\pi},\label{eq:Psi_D}
 \end{equation}
where
\begin{equation}
S_{z}=\int\left[\frac{\mathrm{i}}{2}\left(z^{*}\frac{\mathrm{d}z}{\mathrm{d}t}-z\frac{\mathrm{d}z^{*}}{\mathrm{d}t}\right)-H_{\mathrm{ord}}(z^{*},z)\right]\mathrm{d}t\,\label{eq:Sz}
\end{equation}
denotes the classical action along the trajectory defined with regard to the matrix element
$H_{\mathrm{ord}}(z^{*},z)=\langle z|\hat{H}_{\mathrm{ord}}(\hat{a}^{\dagger},\hat{a})|z\rangle$.
This  represents
the diagonal elements of the ordered Hamiltonian matrix $\hat{H}_{\mathrm{ord}}(\hat{a}^{\dagger},\hat{a})$. In general,
\begin{equation}
\langle z|\hat{H}|z^{\prime}\rangle=\langle z|z^{\prime}\rangle H_{\mathrm{ord}}(z^{*},z^{\prime}),\label{eq:Hord}
\end{equation}
where $|z\rangle$, $|z^{\prime}\rangle$ denotes two arbitrary coherent states.

In coordinate space,
\begin{equation}
\langle x|z\rangle=\left(\frac{\gamma}{\pi}\right)^{1/4}\exp\left[-\frac{\gamma}{2}(x-p)^2+
\mathrm{i}p(x-q)+\frac{\mathrm{i}pq}{2}\right]\label{eq:ansatz}\end{equation}
is a Gaussian wave packet centred at the phase space coordinates $q$ and $p$. 

One should note that, for a static field and a Gaussian potential given by Eq.~(\ref{eq:gaussian}), the effective CCS Hamiltonian is given by 
\begin{equation}
 H_{\mathrm{ord}}(p,q)=\frac{\gamma} {4}+\frac{p^2}{2}-\left(\frac{\gamma} {\gamma+\lambda} \right)^{1/2}\exp[-\eta q^2]+q\mathcal{E}_0,
 \label{eq:Hord2}
\end{equation}
with $\eta=(\lambda\gamma/(\gamma+\lambda))$. Physically, averaging the Hamiltonian over a Gaussian coherent-state basis effectively lowers the potential barrier by introducing an effective, shallower potential. In comparison with the classical Hamiltonian, Eq.~(\ref{eq:Hord2}) shows an effective energy shift $\gamma/4$ \cite{Miller2002,Child2003}. 
\subsection{Phase space configurations and bifurcations} 
\label{sec:config}

\subsubsection{Identifying bound and continuum regions}

\begin{figure*}[tbp]
\begin{minipage}{0.5\textwidth}
\centering
\includegraphics[width=\textwidth]{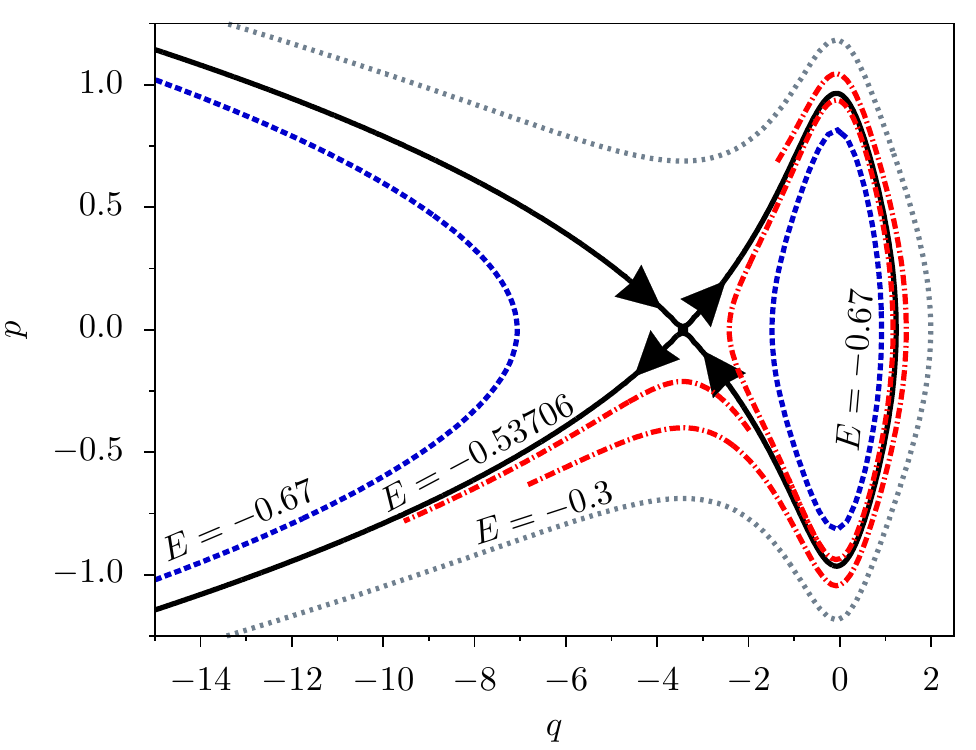}
\end{minipage}
\begin{minipage}{0.5\textwidth}
\centering
\includegraphics[width=\textwidth]{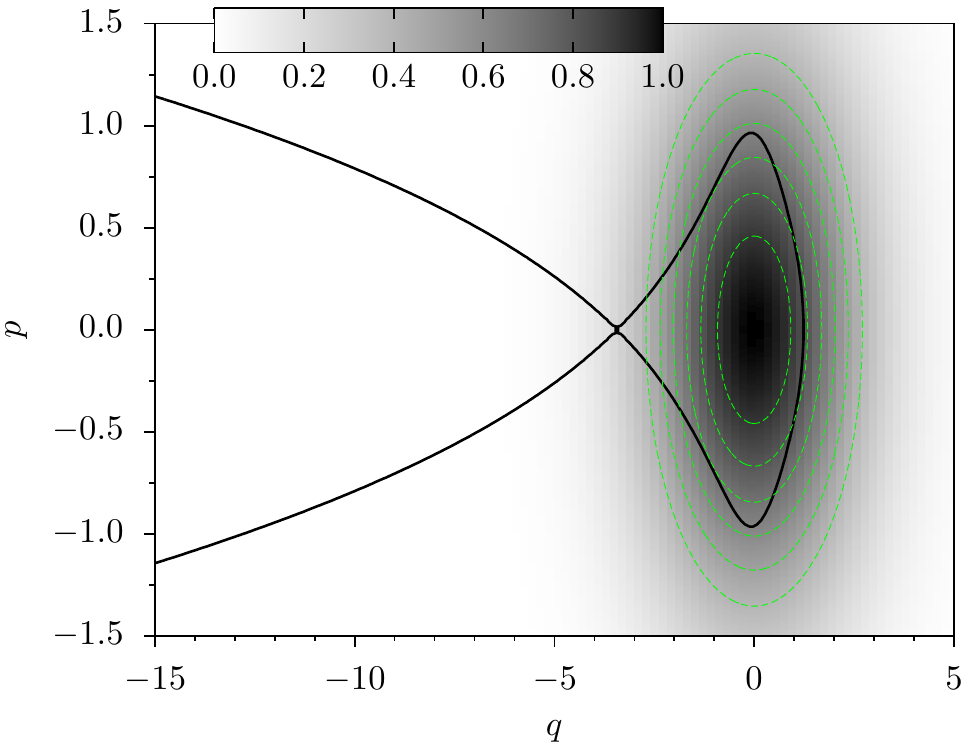}
\end{minipage}\\
\begin{minipage}{0.5\textwidth}
\centering
\includegraphics[width=\textwidth]{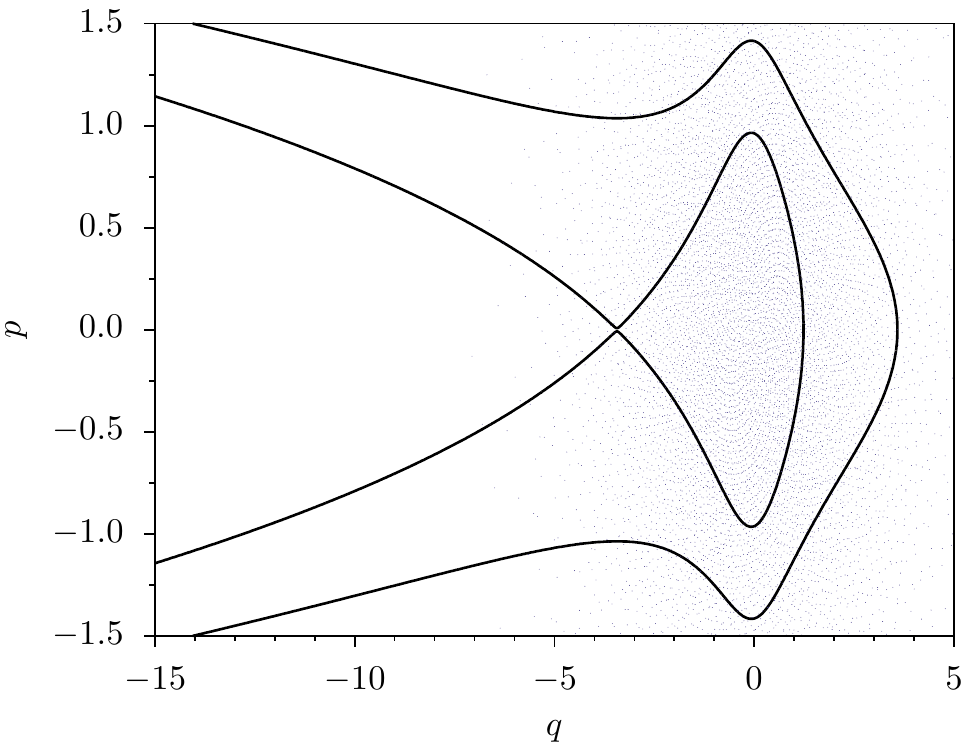}
\end{minipage}
\begin{minipage}{0.5\textwidth}
\centering
\includegraphics[width=\textwidth]{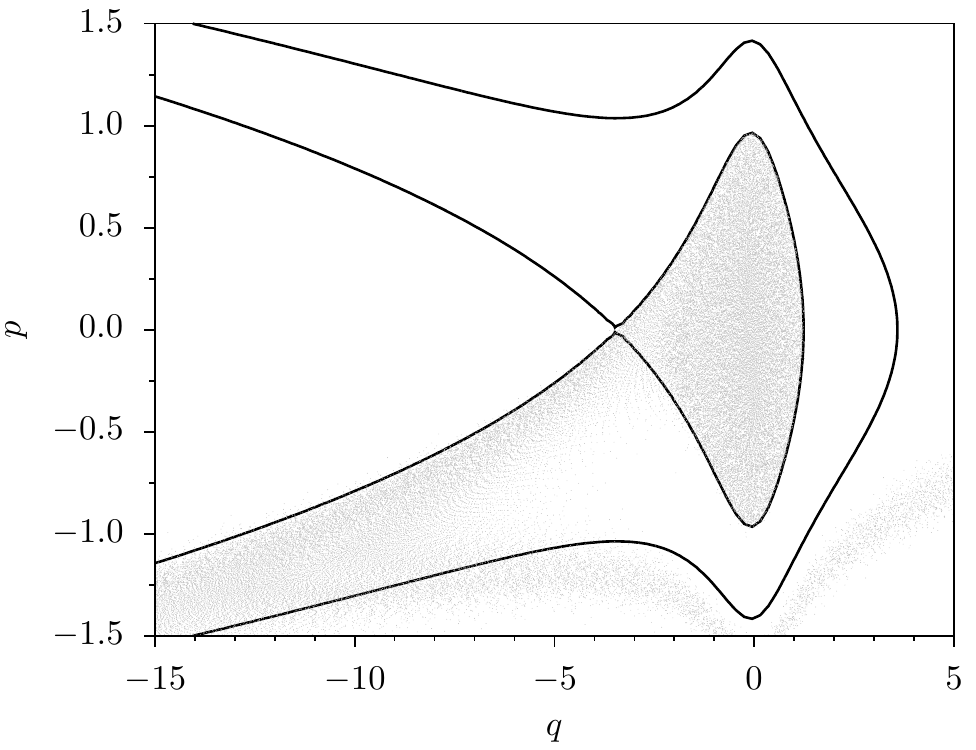}
\end{minipage}
\caption{Upper left panel: Phase portrait of the system defined as an atom with a soft-core potential in a static field of amplitude $\mathcal{E}_0=0.075$ a.u.  ($I=1.97\times 10^{14} \mathrm{W}/\mathrm{cm}^2$). 
Solid lines depict the separatrix in phase space, while dashed and dotted lines illustrate solutions for energies $E=-0.67$ a.u. ($E=-18.23$ eV) and $E=-0.3$ a.u. ($E=-8.16$ eV),
respectively. The dashed-dotted lines show the evolution of sample trajectories from $t=0$ to $t=20$ a.u. Upper right panel: phase space spread for an initial Gaussian wave packet centred at the origin and width $\gamma = 0.5$ (bound-state energy $E \simeq −0.67$ a.u. ($E \simeq -18.23$ eV). Lower left panel: phase space positions of the classical trajectory ensemble mimicking the initial wave packet in the upper right panel. Lower right panel: phase space positions of a classical-trajectory distribution whose initial conditions are shown in the lower left panel, after a time $t=20$ a.u. of static-field propagation. The color bar on the top right panel gives the phase space probability density. From \cite{Zagoya2014}.}
\label{fig:separatrix}
\end{figure*}

 A powerful reason to study phase space dynamics in attoscience is that trajectories can easily be understood as bound or unbound depending on the phase space configuration. This is crucial when looking at signatures of strong field tunneling and over-the-barrier ionisation. Bound and continuum regions can be defined by inspecting the phase portrait of the system. This is exemplified with Fig.~\ref{fig:separatrix}, in which we consider a model atom in a static field. Similar pictures have been provided in \cite{Heller1987,Czirjak2000}. The figure shows two fixed points according to the definition provided in Sec.~\ref{sec:classicalphspc}: a centre at the origin,  and the Stark saddle to the left, whose positions are determined by the minimum and the maximum of the effective potential $V_{\mathrm{eff}}(x)$, respectively. The figure also shows a separatrix,  which corresponds to the stable and unstable manifolds of the Stark saddle. It is associated with the minimum energy to undergo over-the-barrier ionisation. The closed region to the right of the saddle is bound: trajectories within it will propagate along closed orbits. If the particle starts on the left of the Stark saddle, or if it has an energy higher than that of the separatrix, it will be free. This clearly shows that the particle's energy is paramount to defining the continuum regions; the particle being close to the core is not sufficient for describing its dynamics.

 Quantum mechanically, there will always be an uncertainty for the initial wave packet, which will manifest itself as a phase space spread (see Fig.~\ref{fig:separatrix}, upper right panel). In trajectory-based grid methods, such as IVRs, one may employ an ensemble of classical trajectories to mimic the initial spread (see Fig.~\ref{fig:separatrix}, lower left panel). One should note that, due to the initial uncertainty, some trajectories belonging to the ensemble will be in the continuum from the start. Although they cannot cross separatrices, they do from a tail whose energy is above that of the saddle. Energies above the saddle indicate above-the-barrier ionisation (see Fig.~\ref{fig:separatrix}, lower right panel) \cite{Zagoya2014}. 
 
 For a time-dependent field of the form (\ref{eq:efield}), in the parameter range considered in this work, namely low frequencies and high driving-field intensities, one may assume that the system behaves quasi-statically. This means that one may use the approximation that the phase space configurations discussed for static fields and shown in Fig.~\ref{fig:separatrix} hold for each instant of time. They will change instantaneously, such that an electron reaching the continuum at different times and propagating in the field will be exposed to a wide range of transient bound and continuum regions. This implies that the time should be included as an additional variable in an extended phase space. 
 
\begin{figure}
     \centering
     \includegraphics[width=7cm]{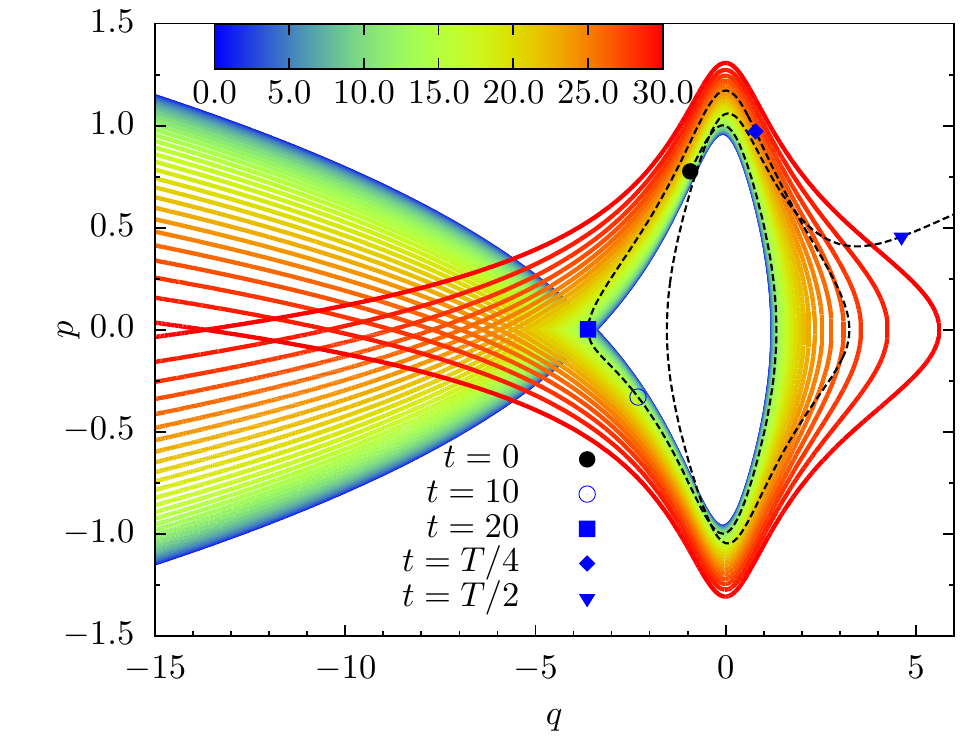}
     \caption{Time-dependent separatrix for half a cycle of the laser field given by Eq.~(\ref{eq:efield}) with $\phi=0$ (solid lines), amplitude $\mathcal{E}_0=0.075$ a.u. ($I=1.97\times 10^{14} \mathrm{W}/\mathrm{cm}^2$) and frequency $\omega=0.05$ a.u. ($\lambda=911 ~\mathrm{nm}$), together with an initially unbound electron trajectory (dashed line), whose initial conditions are indicated by a dot. The gradient color indicates the time variation from $t=0$ (blue) to $t=30$ a.u. (red), which spans slightly less than the first quarter cycle of the the field. The remaining symbols give the phase space coordinates of the particle at specific times $t$. For $t>T/4$, the bound phase space region starts to decrease, until the trapped trajectory is eventually able to escape. The color bar on the top of the figure indicates how the separatrix varies with time, from $t=0$ up to $t=30$ a.u. From \cite{Zagoya2014}.}
     \label{fig:TDseparatrix}
 \end{figure}

 The Stark saddles will be symmetric with regard to the origin for consecutive half cycles. An illustration of a time-dependent separatrix for a field given by Eq.~(\ref{eq:efield}) is displayed in Fig.~\ref{fig:TDseparatrix}. The bound region enclosed by the separatrix increases with time in the first quarter cycle of the field, as it evolves from a maximum to a crossing. This evolution is indicated up to $t=30$ a.u. (see upper color bar), which is less than a quarter of a cycle. Thereafter, sign of the field will reverse and the bound region will decrease until the minimum of the field, in the subsequent half cycle, is reached.
 
 The picture also shows a sample phase space trajectory, in which an electron performs two revolutions around the core until it eventually escapes. This orbit is indicated by the dashed line and should be followed in the clockwise direction, starting from the black dot. The electron is initially in the continuum, that is, outside the region enclosed by the separatrix at $t=0$ (blue curve). It follows this separatrix  closely, but is captured as the bound region increases. At $t=20$ a.u. (solid square), it is in a bound region far away from the saddle indicated by the dark yellow curve. This region increases up to a quarter of a cycle. Thereafter, the oscillating field reverses its sign, and the electron performs a second revolution around the core, which, in phase space, is roughly the mirror image of the first with regard to the momentum axis. Eventually, the electron escapes as it is once more in the continuum (blue triangle). 
\begin{figure}
     \centering
    \includegraphics[width=11cm]{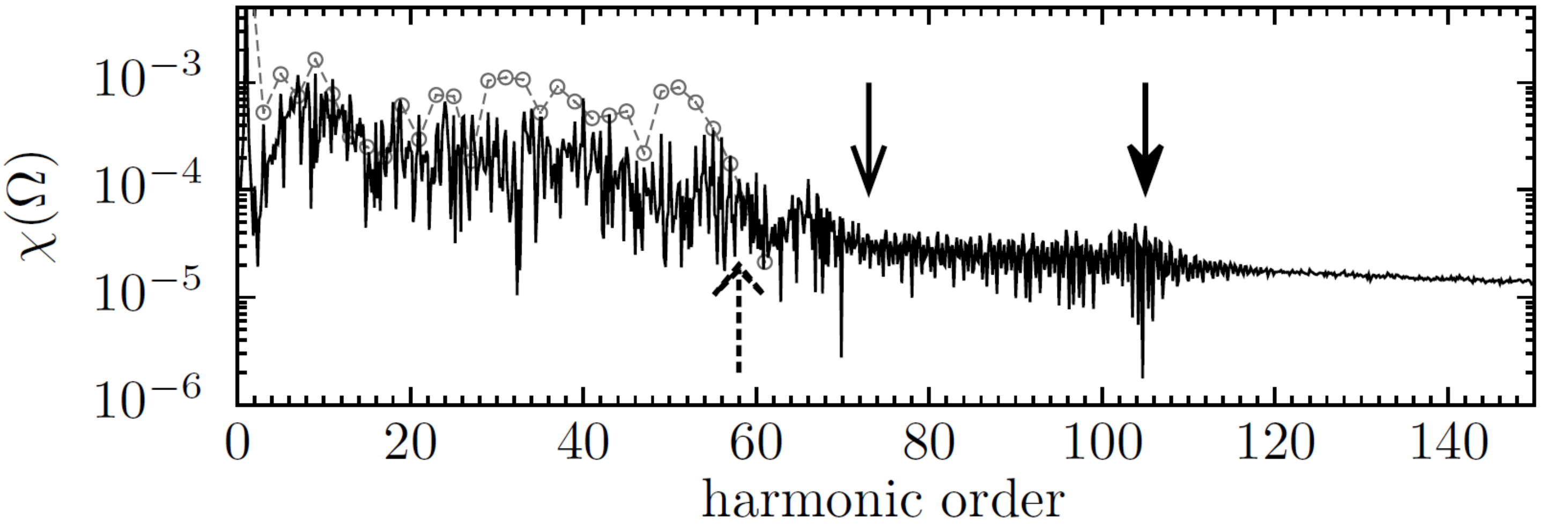}
     \caption{HHG spectra computed from the dipole acceleration (\ref{eq:acceleration}) obtained with inhomogeneity parameters \( \beta\) = 0.004 (solid line) and for the homogeneous case (gray dots). The external field is given by Eq. (\ref{eq:inhom}), and its temporal part by Eq. (\ref{eq:efield}),  with frequency $\omega$ = 0.05 a.u. ($\lambda=911 ~\mathrm{nm}$), amplitude $\mathcal{E}_0= 0.075$ a.u. ($I=1.97\times10^{14} \mathrm{W}/\mathrm{cm}^2$) and phase \(\phi = \pi/2.\) The pulse duration is 6 cycles. The cutoff harmonics are indicated by the arrows in the figure. From \cite{Zagoya2016}}
     \label{fig:figinhomogeneousHHG}
 \end{figure}
\begin{figure*}
\begin{minipage}{\textwidth}
\centering
\includegraphics[width=14cm]{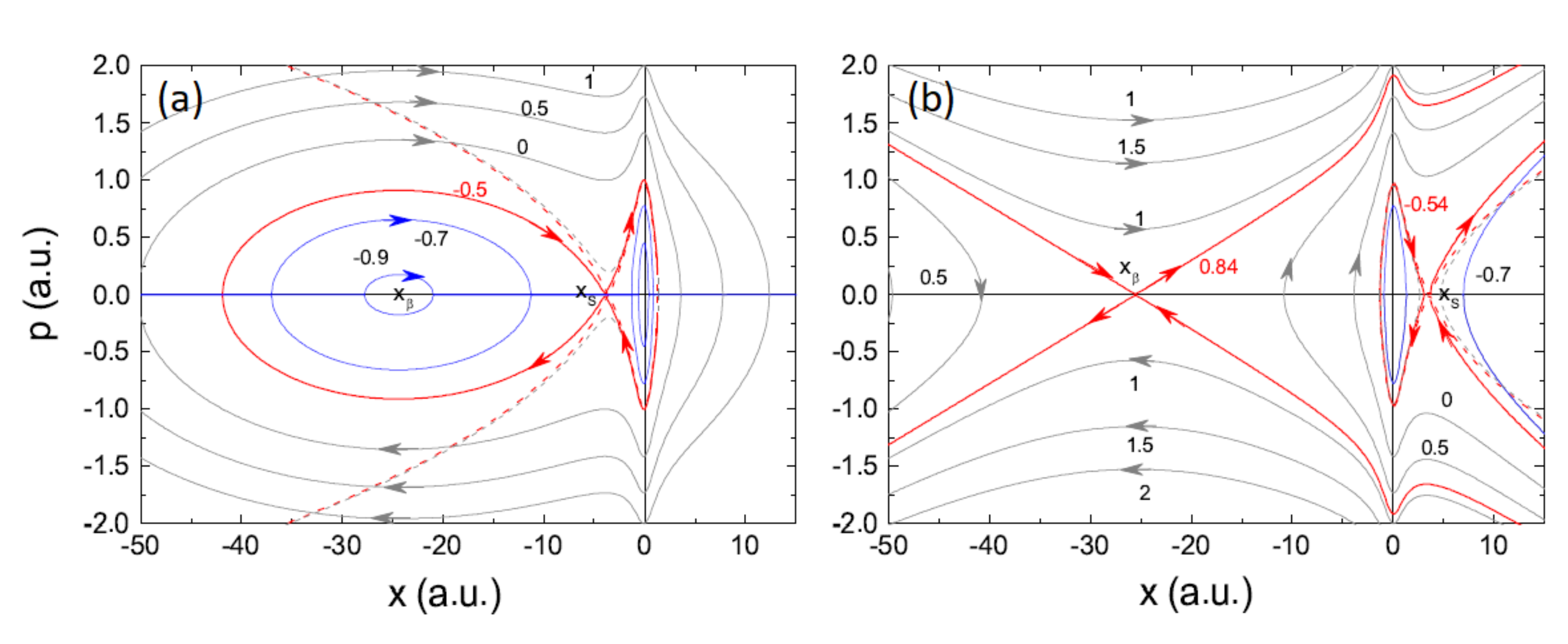}
\end{minipage}\\
\begin{minipage}{\textwidth}
\centering
\includegraphics[width=12cm]{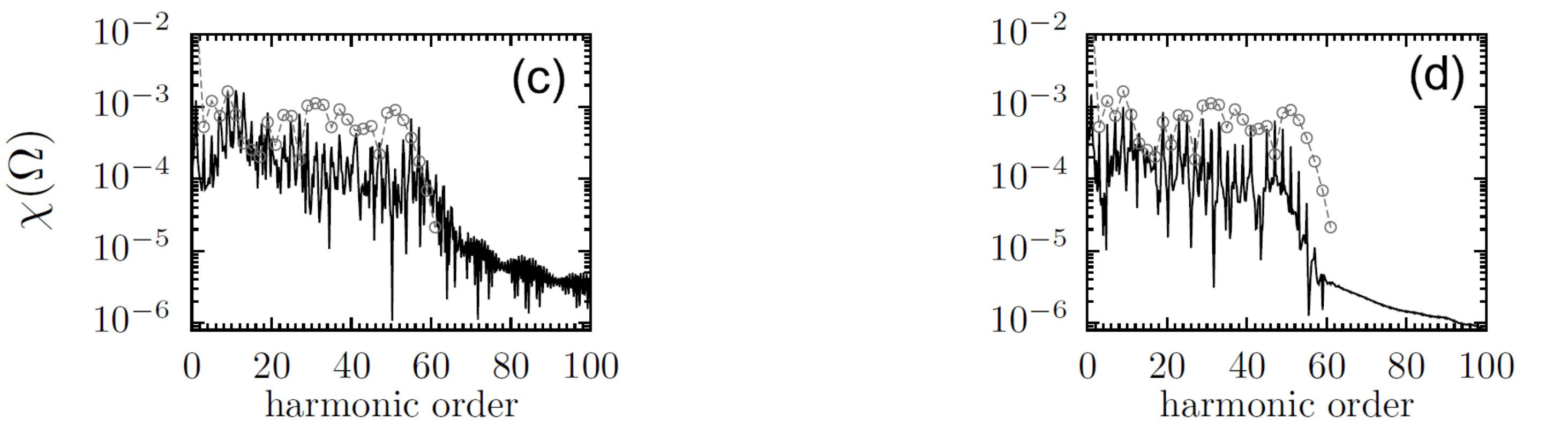}
\end{minipage}
\caption{ [Upper row] Phase portraits calculated for the Hamiltonian (\ref{eq:Hoperator}), with soft core potential (\ref{eq:softcore}) and inhomogeneous field (\ref{eq:inhom}) with $\beta = 0.04$. We consider the field to be (a) $\mathcal{E}_0 = 0.07$ a.u. ($I=1.72\times10^{14} \mathrm{W}/\mathrm{cm}^2$) or (b) $\mathcal{E}_0 = -0.07$ a.u. ($I=1.72\times10^{14} \mathrm{W}/\mathrm{cm}^2$). The separatrices are given by the red lines in the figure, and the numbers near each contour denote the corresponding total energy of the system. The Stark saddle $x_S$ and the fixed point $x_\beta$ due to the inhomogeneity are indicated in the figure. The contours in blue are related to the energies lower than that of the Stark saddle. The red dashed lines give the separatrices for the homogeneous case $\beta = 0$, which occur at at energy $E_{sep} =-0.52 $ a.u. ($E_{sep} = -14.15$ eV). The black dashed lines give the phase space trajectory for energy $E = -0.5$ a.u. ($E = -13.61$ eV) and $\beta = 0$.
[Lower row] High-order harmonic spectra computed using the dipole acceleration (\ref{eq:acceleration}) for the same parameters as Fig.~\ref{fig:figinhomogeneousHHG} using two symmetric toy models which we artificially constructed from \cite{Zagoya2016}. Panel (c) corresponds to the two-saddle potential and panel (d) to the two-centre potential. The gray dots give the spectra computed for the homogeneous case. From \cite{Zagoya2016}.}
\label{fig:concaveConvex}
\end{figure*}

\subsubsection{Inhomogeneous fields and phase space configurations}
 
 Time-dependent separatrices and transient phase space configurations play a key role for inhomogeneous fields, which are employed to model HHG in alternative media such as nanostructures \cite{Zagoya2016}. The laser field becomes enhanced by plasmonic resonances and the external field has to be regarded as inhomogeneous. Whilst it is well known in the literature that even small inhomogeneities may lead to huge changes in the high-order harmonic spectra \cite{ciappina2012,Ciappina2012b,shaaran2012,Yavuz2012,shaaran2013,Hekmatara2014,Luo2014}, such as an increased cutoff energy (see Fig.~\ref{fig:figinhomogeneousHHG}), the analysis of phase space regions provides further insight.
 
 An inhomogeneous field approximated by Eq.~(\ref{eq:inhom}) introduces an additional fixed point with regard to its homogeneous counterpart, whose nature changes with the sign of the laser field. For $\mathcal{E}(t)>0$ or $\mathcal{E}(t)<0$, where $\mathcal{E}(t)$ is given by Eq.~(\ref{eq:efield}), it will be a centre or a saddle, respectively.
The two configurations introduce either a concavity, see Fig.~\ref{fig:concaveConvex}(a),(c) or a convexity, see Fig.~\ref{fig:concaveConvex}(b), (d), to the dynamical system and are a crucial tool in understanding the increase in the HHG cutoff energy, previously attributed to the higher electron momentum for the convex system at the instant of ionisation \cite{shaaran2012}.

However, by separating our configurations into two toy potentials (one always convex, one always concave), we show that, instead, the increased cutoff energies are due to the concavity providing additional confinement and forcing high-energy orbits back to the core. This is a similar mechanism to that in \cite{Faria2000}, in which an additional confining potential led to a substantial increase in the cutoff energy without loss of intensity. Indeed, Fig.~\ref{fig:concaveConvex}(c),(d), shows that only the spectrum obtained from a concave potential leads to an increase in cutoff energy.

\subsubsection{Enhanced ionisation and nested separatrices}
\begin{figure}[ht]
    \centering
    \includegraphics[width=9cm]{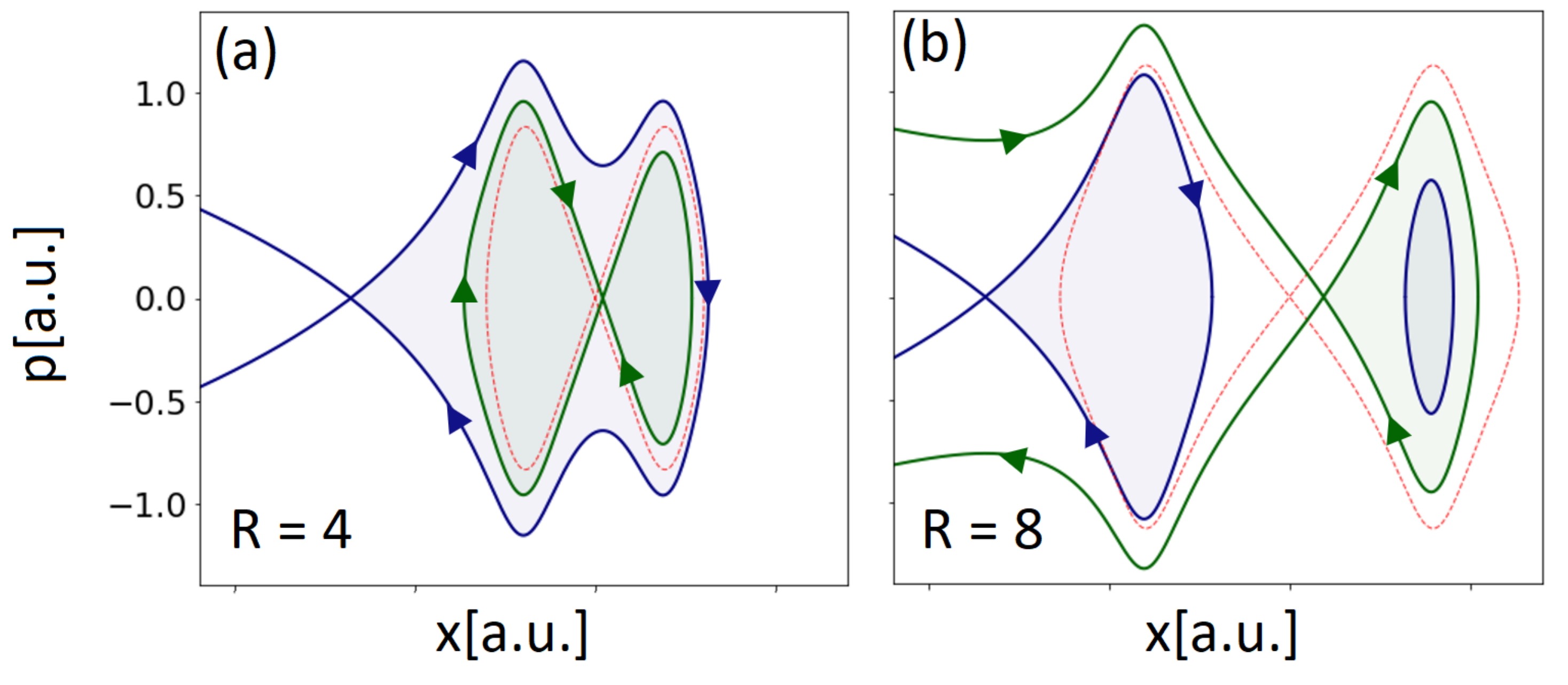}
    \includegraphics[width=9cm]{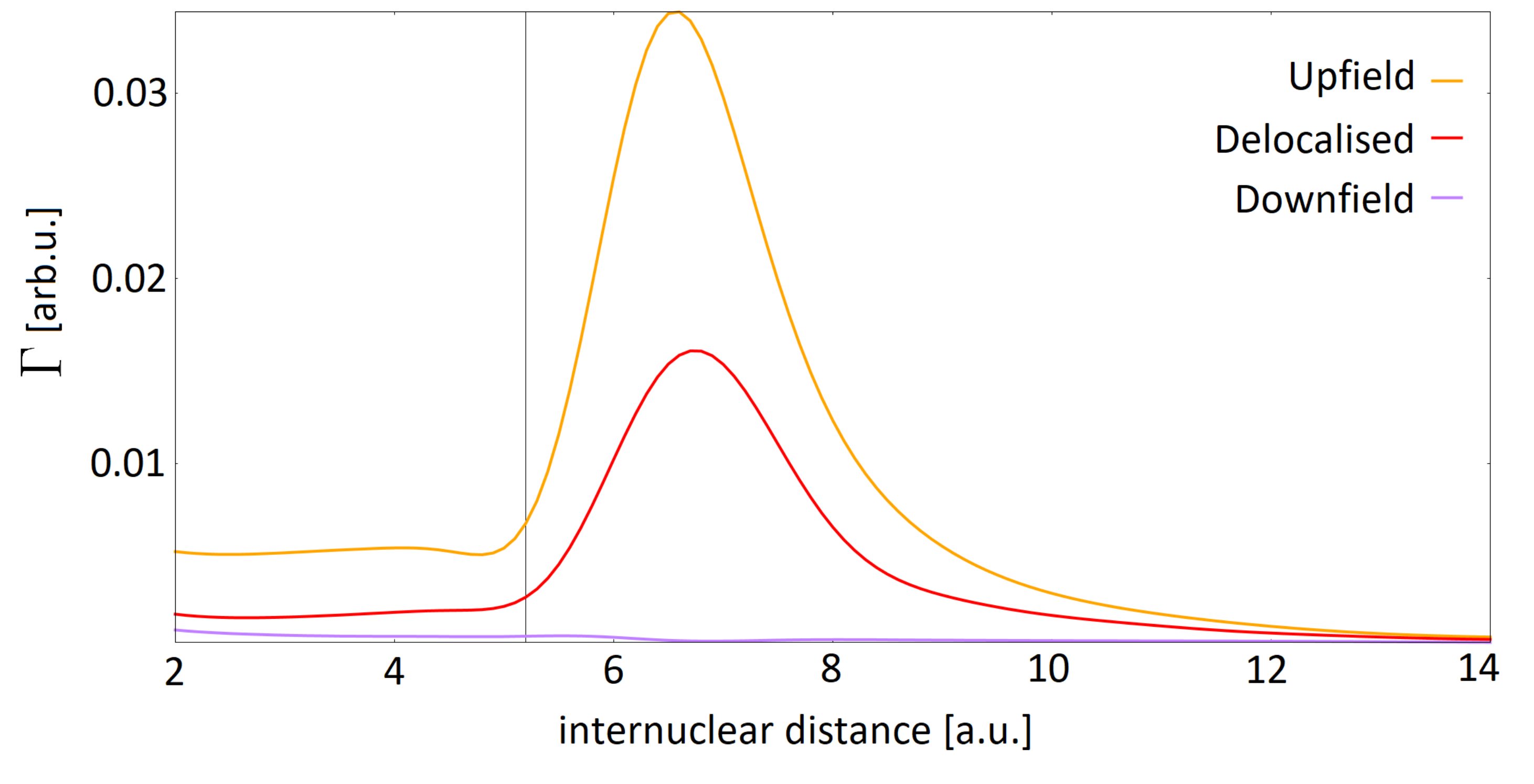}
    \caption{[Upper row] Phase portraits for the one-dimensional homonuclear molecular models described by the binding potential Eq.(\ref{eq:molpot}), using inter-nuclear separations in panel (a) of $R = 4$ a.u. and in panel (b) $R = 8$ a.u. with a static field $\mathcal{E}_0 = 0.0534.$ a.u ($I=10^{14} \mathrm{W}/\mathrm{cm}^2$). The field-free separatrices and potentials are given by the dashed red lines. The shaded areas indicate the phase space regions for which the wave packet is bound. The colours of these regions match those of the respective separatrices. [Lower row] Ionisation rate~(\ref{eq:ionisation}) as a function of the inter-nuclear distance R, using different starting wave packets: delocalised (red), localised upfield (orange) and localised downfield (purple). The vertical line indicates the inter-nuclear separation for which the phase space configuration changes. From \cite{Chomet2019}.}
    \label{fig:bif2}
\end{figure}

The change in the nature of the fixed points due to the inhomogeneous field leads to different configurations, depending on the external laser field. This is a clear case of a bifurcation. Studying the $\mathrm{H}_2^+$ molecular system in \cite{Chomet2019} using phase space diagrams also leads to several fixed points, shown in Fig.~\ref{fig:bif2}. Thereby, one may identify a Stark saddle to the left, a central saddle between the two molecular wells and two molecular centres separated by $R$. We again have two different configurations, not due to a time dependent field, but to a bifurcation for increasing internuclear separation $R$.

 For $R$ below the critical (bifurcation) value, see Fig.~\ref{fig:bif2}(a), the separatrices are nested and classical trajectories close to the core stay bounded. If the energy of the central saddle becomes greater than that of the Stark saddle, the separatrices open up, see Fig.~\ref{fig:bif2}(b). Then, a trajectory in the upfield molecular well only needs to pass the upfield potential barrier in order to reach the continuum. This explains the behaviour of the ionisation rate with respect to the internuclear distance, shown in Fig.~\ref{fig:bif2}(c). Indeed, after the bifurcation and as the separatrices open up there is a sharp increase in the ionisation rate. Additionaly, the ionisation rate for a wave packet localised around the upfield molecular well, Eq.~(\ref{eq:Psi0}), is about double that of a delocalised wave packet, Eq.~(\ref{eq:statcat}), and the ionisation rate for a wave packet localised downfield, Eq.~(\ref{eq:Psi0}), is suppressed. From this it is deduced that maximum enhancement is achieved if the energy of the saddle is high enough for the tunnelling electron coming from the upfield population not to be trapped by the downfield centre, but low enough for the effective potential barrier to be narrower than that of a single atom.

\subsection{Semi-classical versus quantum regimes}
\label{sec:clvsquantum}

In the following, we will illustrate different physical regimes, which may or may not have a classical counterpart. Thereby, we will also exemplify roles that trajectory-ensembles may play. Classical-trajectory ensembles allow for an intuitive understanding of an electronic wave packet's evolution when paired with quantum mechanical methods. They may, for instance, be employed to link the outcome of quantum-mechanical computations, such as the TDSE, to the picture of a returning electron, or be used to construct grids for initial-value representations. 

\subsubsection{Pairing trajectories with quantum mechanical methods}
\label{subsub:semicallical}
\begin{figure}[ht]
    \centering
    \includegraphics[width=11cm]{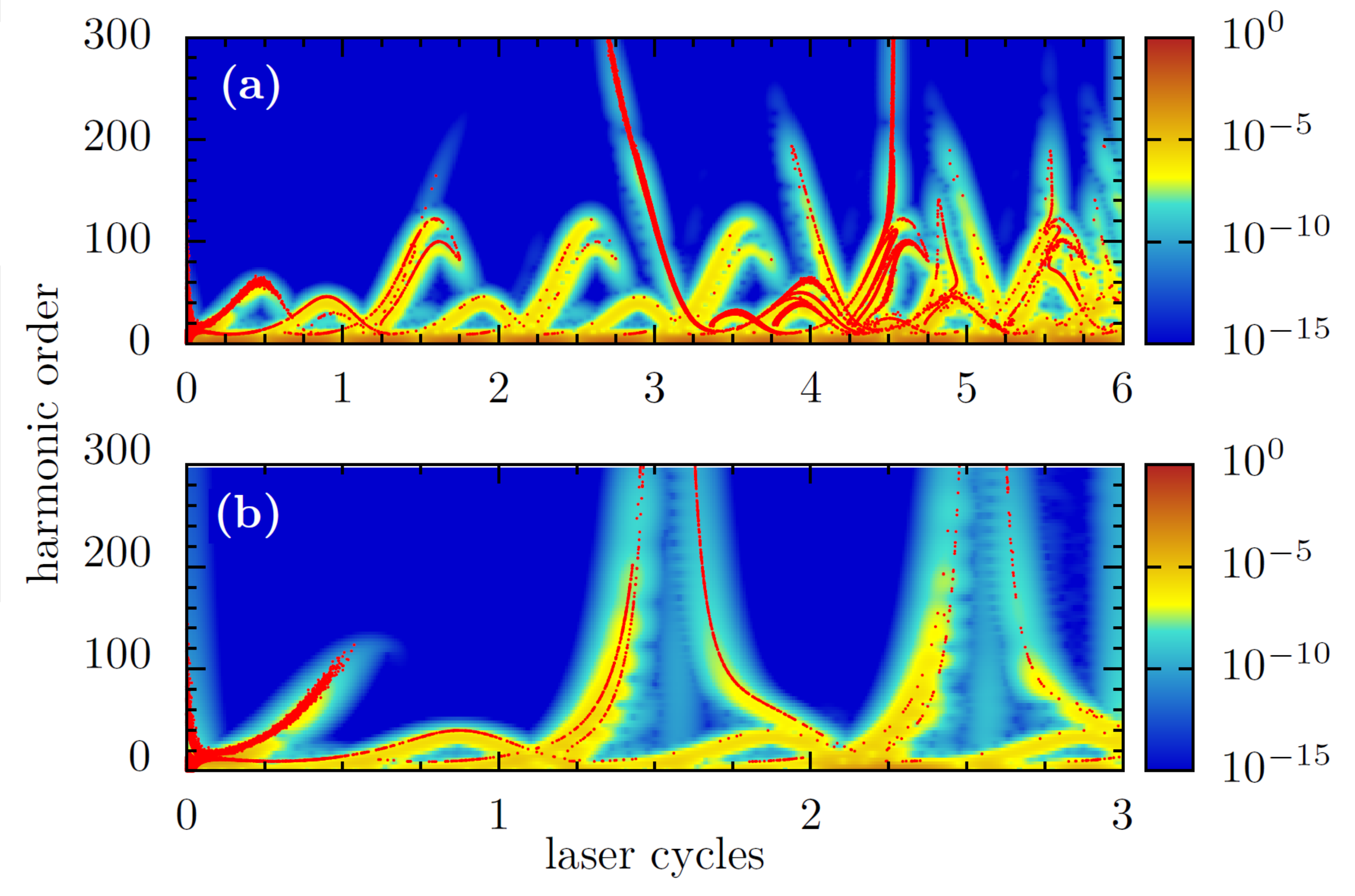}
    \caption{Time-frequency maps computed from the dipole acceleration, and classical returning times (superimposed dots) as functions of the harmonic order and the field cycles with frequency \(\omega\) = 0.05 a.u. ($\lambda=911 \hspace*{0.1cm}\mathrm{nm}$), amplitude $\mathcal{E}_0= 0.075$ a.u. ($I=1.97\times10^{14} \mathrm{W}/\mathrm{cm}^2$) and phase \(\phi = \pi /2.\) The pulse duration is 6 cycles. The inhomogeneity parameter is (a) \( \beta\) = 0.01 and (b) \( \beta\) = 0.02. The color bars indicate the intensity of the time-frequency signal. From \cite{Zagoya2016}.}
    \label{fig:quant0}
\end{figure}

A widespread example is to use trajectories to infer electron return times from HHG spectra.  This is known since the 1990s \cite{Antoine1996}, and the fact that these times are well specified within a field cycle has paved the ground for attosecond-pulse generation. It is also well known that windowed Fourier transforms of the time-dependent dipole acceleration [Eq.(\ref{eq:acceleration})] lead to periodic arch-like structures which can be explained using classical-trajectory ensembles and give pairs of return times merging at the cutoff \cite{Faria1997,deBohan1998}. 

Some of these structures are illustrated in Fig.~\ref{fig:quant0}, in which the return times of an ensemble of classical trajectories are used to understand the cause of plasmonically enhanced HHG. While this is a quantum-mechanical process, when the trajectories are paired with time-frequency (Gabor) maps computed from the TDSE, they can offer very useful insight. There is very good agreement between the two, and we discern peculiar features, which increase with the inhomogeneity of the field.  First, there is a suppression in the arch-like structures that occurs with the same periodicity as the field and happens for times after each field crossing. This can be explained by the phase space dynamics in Fig.~\ref{fig:concaveConvex}: For times prior to the crossing, the prevalent configuration, with two centres, forces the electron to return, while the phase space configurations subsequent to it, with two saddles, may contribute to irreversible ionisation. This confirms that confinement plays a more important role than the electron reaching the continuum with a higher velocity.  
A second feature are structures occurring over several field cycles that extend up to very high frequencies. These structures are associated with a second time scale introduced by the additional centre, which will be briefly discussed in Sec.~\ref{sub:timescales}. The frequency with which they occur increases with the inhomogeneity parameter $\beta$. Further details are provided in our original publication \cite{Zagoya2016}.

\begin{figure}
    \centering
    \includegraphics[width=9cm]{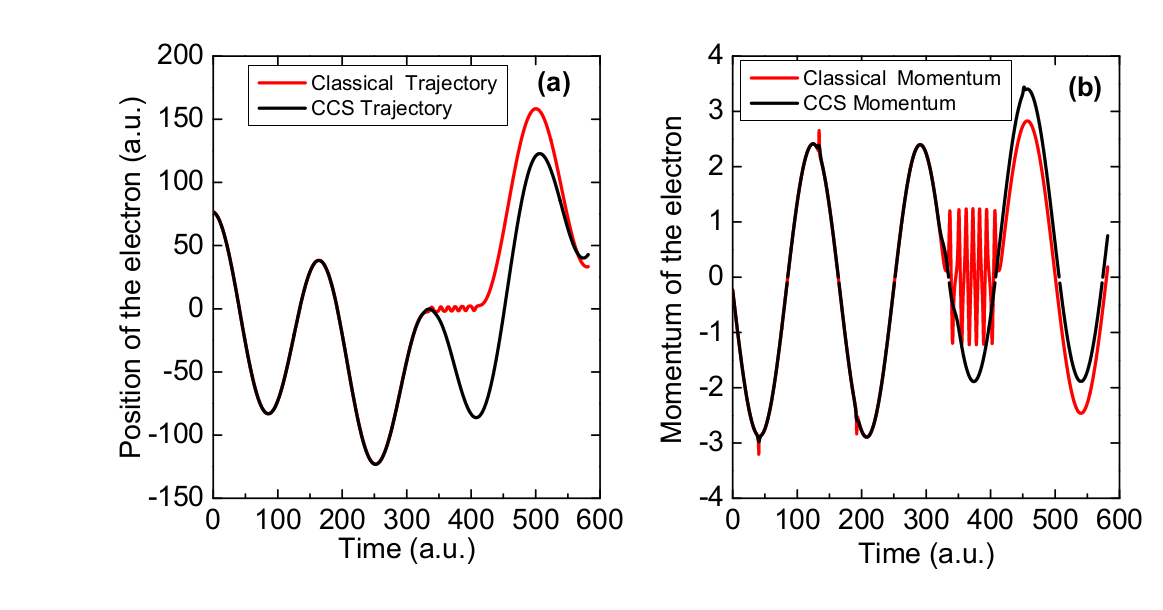}
    \caption{Time-dependent coordinate and momenta computed using the 1D Gaussian potential (\ref{eq:gaussian}) and the Hamiltonians $H_{\mathrm{cl}}(p,q)$ (Eq.~(\ref{eq:Hclassical})) and $H_{\mathrm{ord}}(p,q)$ (Eq.~(\ref{eq:Hord})) (red and black lines, respectively), in an external time dependent field of amplitude $\mathcal{E}_0=0.1$ a.u. ($I=3.51\times 10^{14} \mathrm{W}/\mathrm{cm}^2$) and frequency $\omega=0.0378$ a.u. ($\lambda=1205~\mathrm{nm}$). In panel (a), the initial position has been chosen as $x_0=76.72464472$ a.u., while in panel (b) the initial momentum was taken as $p_0= −0.232201645$ a.u.. From \cite{Wu2014}.
    }
    \label{fig:Jiethesis1.png}
\end{figure}
In addition to that, classical-trajectory ensembles  are employed to construct time-dependent grids which guide initial-value representations (IVRs). In \cite{Zagoya2014}, we assess the suitability of IVRs for modeling tunnel ionisation in an unusual setting, namely starting from a bound state. We perform these studies in phase space using the Wigner quasiprobability distribution constructed from IVRs and from the TDSE.  Therein, we consider the Herman Kluk propagator, which is a semiclassical IVR, and the Coupled Coherent States (CCS) method, which is a quantum IVR solving the TDSE in a coupled coherent state basis. The two Hamiltonians are stated in Sec.~\ref{subsub:IVRs} and differ, because the quantum corrections due to the coherent-state averaging introduce energy shifts and effectively lower the potential barrier. This implies that CCS orbits may cross classical separatrices, while the orbits employed in the HK propagator may not. A detailed discussion of the differences between the two propagators is provided in \cite{Miller2002,Child2003} and in our previous work \cite{Zagoya2014}. Fig.~\ref{fig:Jiethesis1.png} illustrates how specific trajectories in position and momentum space differ when using the two IVRs. An important issue is that the quantum corrections will allow the CCS orbits to cross classical separatrices. This leads to the CCS being more accurate at reproducing tunneling, and there will be a degradation of the outcome of the HK propagator for longer times. This degradation has been observed in our previous work \cite{Zagoya2014} for high-harmonic spectra.

\begin{figure*}[tbp]
\begin{minipage}{4.5cm}
\centering
\includegraphics[width=\textwidth]{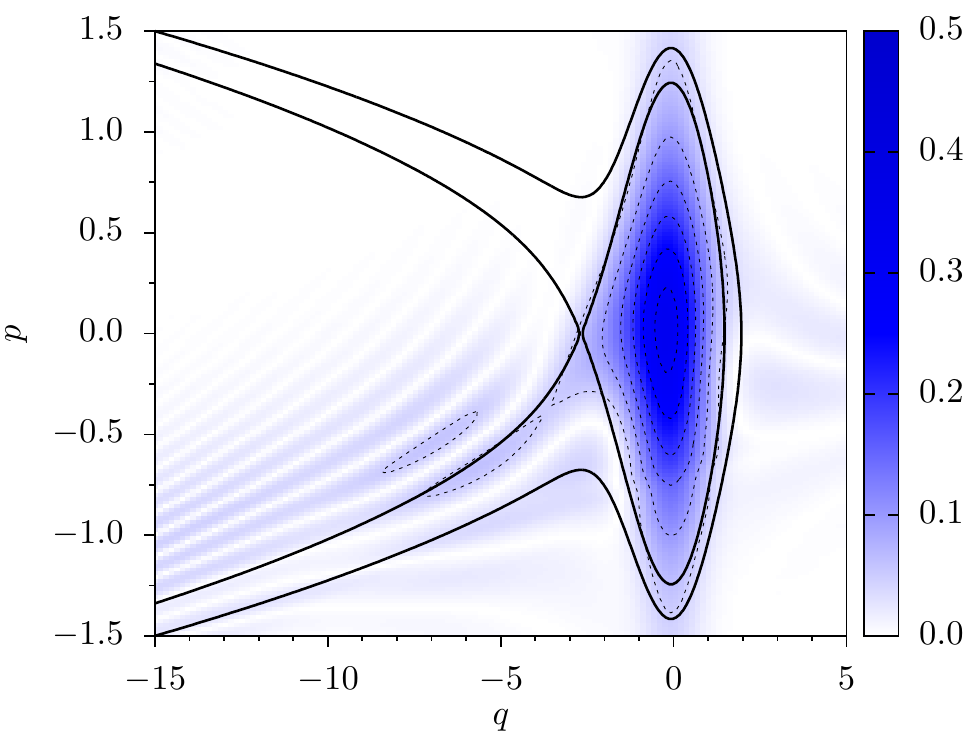}
\end{minipage}\quad
\begin{minipage}{4.5cm}
\centering
\includegraphics[width=\textwidth]{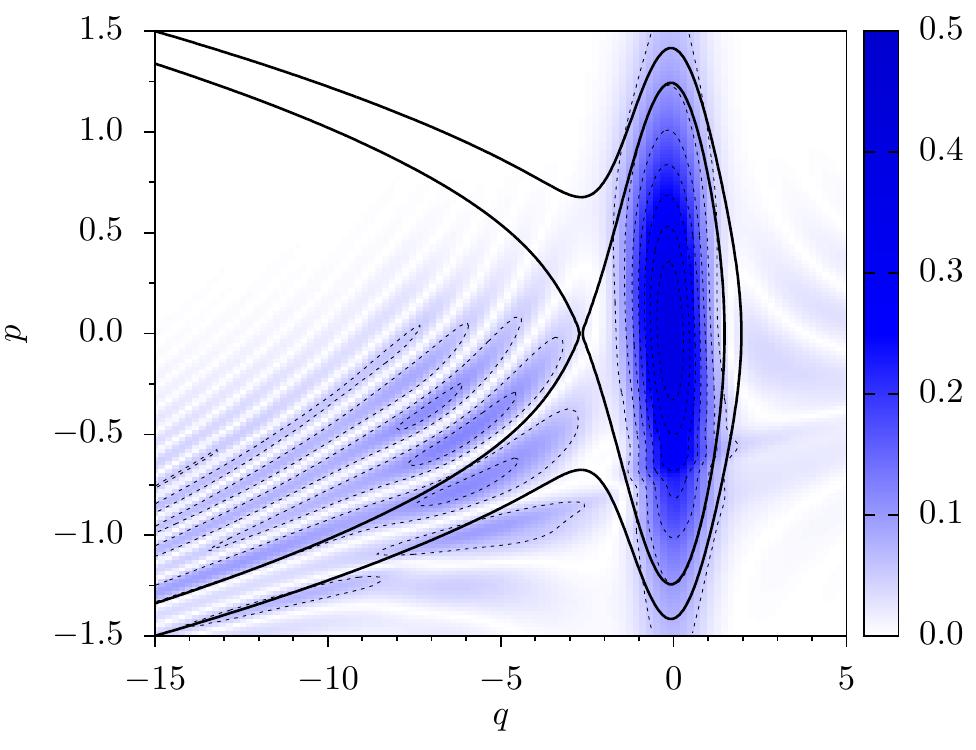}
\end{minipage}\quad
\begin{minipage}{4.5cm}
\centering
\includegraphics[width=\textwidth]{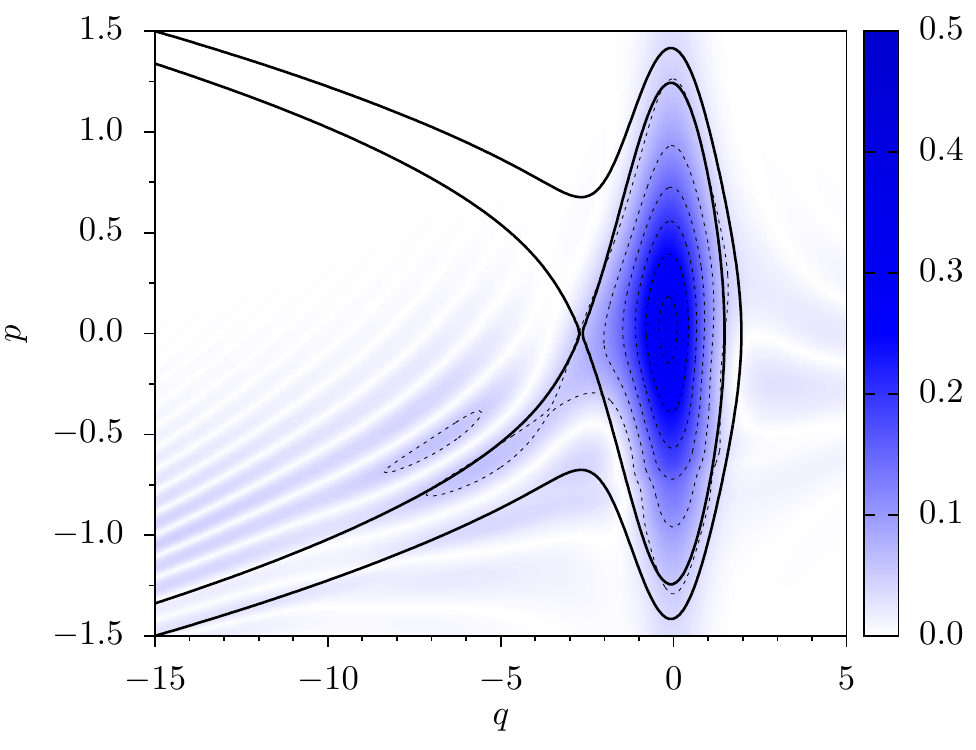}
\end{minipage}
\caption{Modulus squared of the Wigner quasiprobability distributions  computed for a static field of amplitude $\mathcal{E}_0 = 0.075$ a.u. ($I=1.97\times 10^{14} \mathrm{W}/\mathrm{cm}^2$) and the Gaussian potential (\ref{eq:gaussian}), propagated up to $t=20$ a.u.  The left, middle and right panels have been computed with the TDSE, the HK propagator and the CCS method, respectively. The separatrix and the curve in phase space for the energy $E=0$ are illustrated by the thick lines in the figure. For the HK propagator and the CCS method we use $10^{7}$ and
$1600$ trajectories, respectively. The color bars give the square of the Wigner quasiprobability density. From \cite{Zagoya2014}.}
\label{fig:wignergaussian}
\end{figure*}

The agreement between the different methods is illustrated in Fig.~\ref{fig:wignergaussian}, in which Wigner quasiprobability distributions were calculated for a Gaussian potential using the CCS and the HK propagators. In all cases, the Wigner function presents the signature semiclassical tail discussed in \cite{Czirjak2000} (see also \cite{Heller1987} for a seminal discussion of such features in the context of quantum localisation). The tail follows the separatrix and crosses from the bound to continuum region around the Stark saddle. There are also interference fringes on the left-hand side of the saddle, which may be associated with ionisation events in previous times. The agreement between the CCS and the TDSE is much better, with practically identical quasiprobability distributions. Furthermore, significant fewer trajectories are required than in the HK case. This is due to the quantum corrections in the CCS Hamiltonian $H_{\mathrm{ord}}(p,q)$.

Nonetheless, it is remarkable that these quantum features are present despite using HK wave packets and a real-trajectory grid. This is in strong contrast to how the trajectories guiding the grid behave, exemplified in Fig.~\ref{fig:separatrix}: the HK trajectories with over-the-barrier energy never cross phase space barriers, while the Wigner function does. This discrepancy is justified by the non-locality of the Wigner function near separatrices, and by the fact that, near the Stark saddle, the barrier is approximately parabolic \cite{Balazs1990}. This means that, at least for short times, the HK propagator is applicable. One should note, however, that the HK trajectories contribute to forming the tail. Indeed, if the classical trajectories with initial positions outside the bound region are removed, the tail as well the interference fringes are absent \cite{Zagoya2014}. This is consistent with the findings of \cite{Heim2013}, who show that the total weight of the classical phase space trajectories corresponding to energies below, or above the top of a parabolic barrier give the reflection or transmission coefficients.

\begin{figure}[h]
    \centering
    \includegraphics[width=11cm]{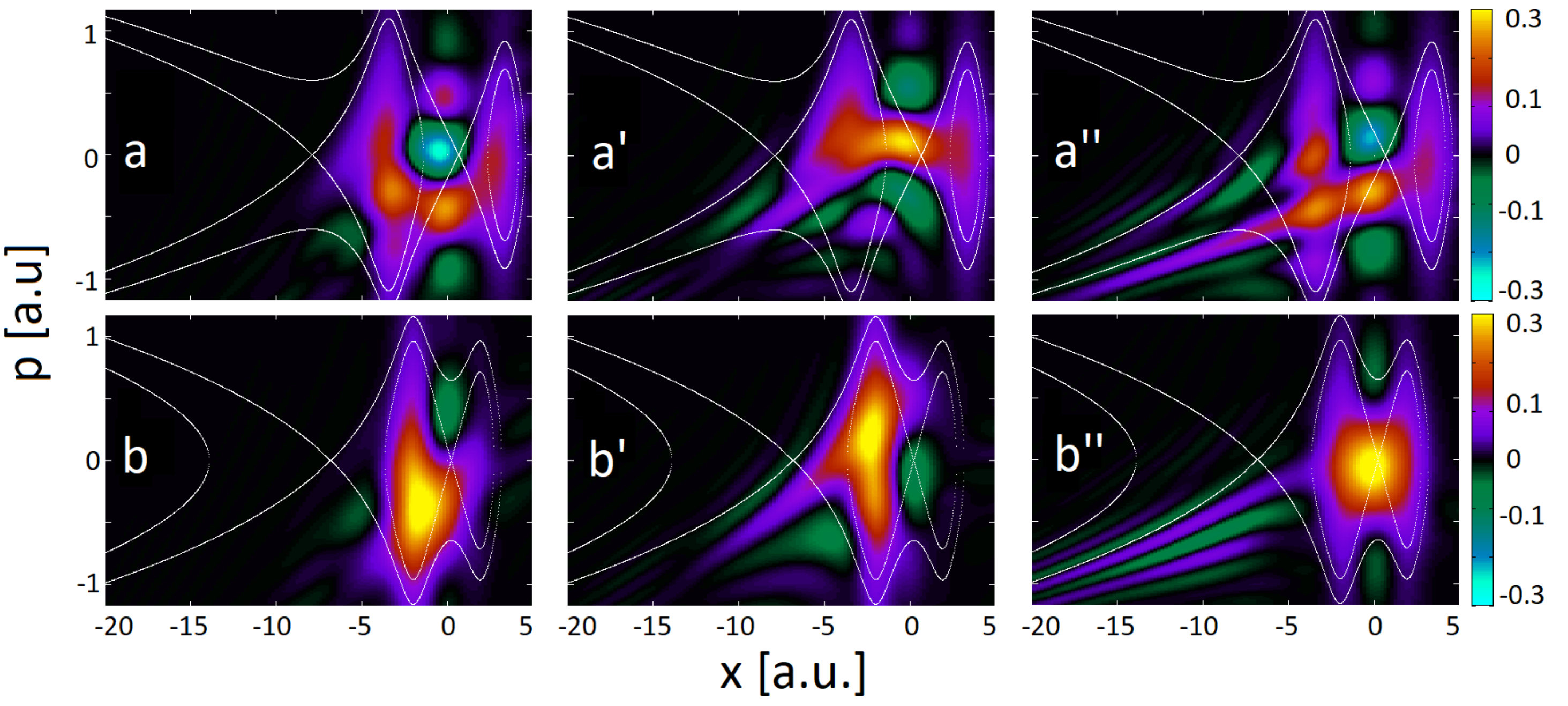}
    \caption{Wigner quasiprobability distribution at different instants of time, calculated for a model $\mathrm{H}_2^+$ molecule in a static field of strength \(\mathcal{E}_0=0.0534\) a.u. ($I=10^{14}  \mathrm{W/cm}^{2}$) using an initially delocalised (cat) state given by Eq.~(\ref{eq:statcat}), with $\gamma=0.5$. In the top and bottom row, the inter-nuclear separation is taken as \(R=6.8\) a.u. and \(R=4\) a.u., respectively. The temporal snapshots are given from left to right. Panels (a) and (a$'$) and (a$'' $) [first row] have been calculated for $t = 8$ a.u., $t = 16$ a.u. and $t = 24$ a.u., respectively. Panels (b), (b$'$) and (b$'' $) [second row]  for $t = 8$ a.u., $t = 16$ a.u. and $t = 30$ a.u.. The thin white lines in the figure give the equienergy curves (including the separatrices). The color bars give the Wigner quasiprobability density. From \cite{Chomet2019}.}
    \label{fig:quantChomet}
\end{figure}
\begin{figure}[ht]
    \centering
    \includegraphics[width=15cm]{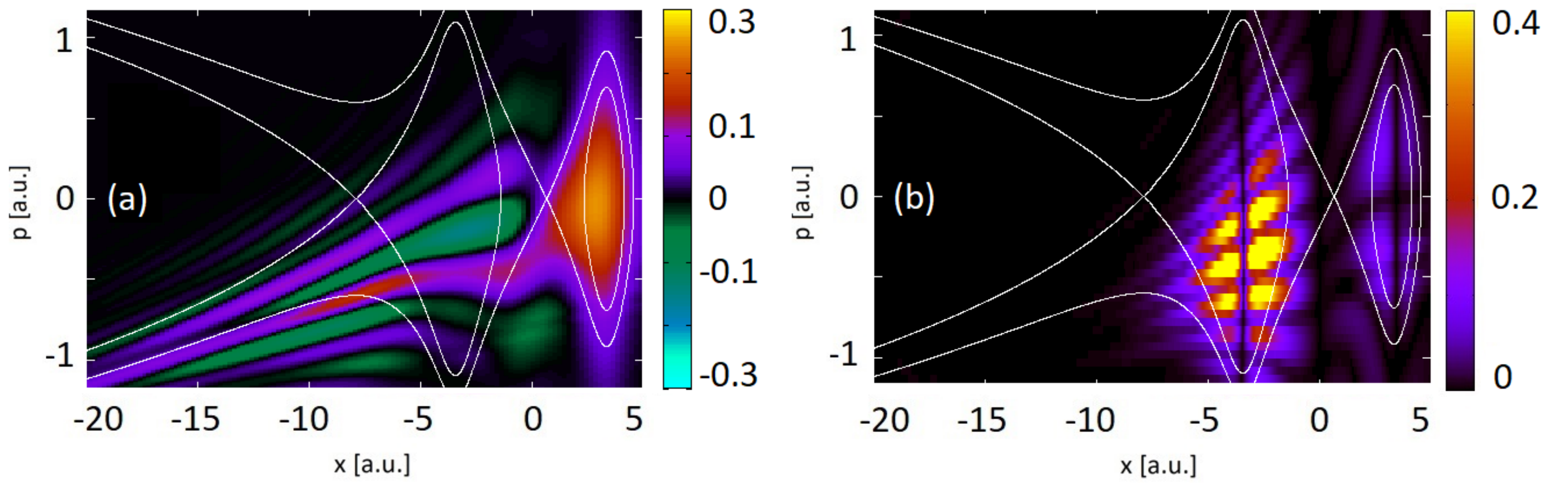}
    \caption{Comparison of (a) the Wigner quasi-probability distribution and (b) the quantum corrections Q(x, p, t), calculated for a model $\mathrm{H}_2^+$ molecule of inter-nuclear separation \(R = 6.8\) a.u. at time \(t = 24\) a.u. in a static field of strength \(\mathcal{E}_0 = 0.0534\) a.u. (\(I=10^{14} \mathrm{W/cm}^{2}\)) using a Gaussian initial wave packet centred around the upfield potential well. The thin white lines in the figure give the equienergy curves (including the separatrices). The color bar on the left panel give the Wigner probability density, while that on the right panel give the magnitude of the quantum corrections. From \cite{Chomet2019}.}
    \label{fig:quant2}
\end{figure}

\subsubsection{Non-classical behaviour in phase space}
\label{subsub:quantum}

In the examples provided above, one clearly sees a classical-quantum correlation, either in structures from time-frequency maps being related to classical return times, or in Wigner function tails following classical separatrices. However, what about situations in which the quantum phase space evolution has no classical counterpart? Below we illustrate the difference between semiclassical and quantum pathways, in the context of molecular enhanced ionisation \cite{Chomet2019}.

The Wigner function nicely illustrates the physical mechanisms behind molecular enhanced ionisation, and these different types of behavior. From this quasiprobability distribution analysis in Fig.~\ref{fig:quantChomet}(a$''$),(b$''$), the signature semiclassical escape tail associated with over-the barrier ionisation is identified, as well as the oscillatory behaviour around the classical separatrix. However, intriguing structures that cycle through the momentum space,  whose behavior does not follow separatrices, are also present. In Fig.~\ref{fig:quantChomet}(a)(b), the quasiprobability distribution starts a cycle from the upfield centre to the downfield well. Using the interference fringes around the central saddle as a quantum bridge, this flow does not follow any of the classical separatrices. As shown in Fig.~\ref{fig:quantChomet}(a$'$),(b$'$), the cyclical movement then transfers part of the downfield population back to the upfield molecular well, while the semiclassical tail starts to form.  In \cite{Takemo2011,Takemoto2010}, those structures were called momentum gates and  were related to the non-adiabatic following of the time dependent field. However, as is seen in Fig.~\ref{fig:quantChomet}, these are present even when using a static field, and for various internuclear distances (before and after the bifurcation in Fig.~\ref{fig:bif2}). This provides evidence that the actual mechanism is intrinsic to the molecule, instead of a response to an external field. In \cite{Chomet2019}, we show that quantum interference builds a bridge between the two centres in the molecule, which supports this direct intra-molecular population flow. For that reason, we refer to the mechanism behind the momentum gates as ``quantum bridges". 

From studying the effect of different internuclear distances as well as different initial wave packet configurations, the optimal configuration for a static field stems from using a localised upfield initial wave packet, see Fig.~\ref{fig:quant2}(a). Indeed, the cyclical motion is absent, meaning the subsequent quantum bridge bringing the population back to the upfield well does not form. Therefore, the initial quantum bridge forms, leading the upfield population through the downfield centre, to the semiclassical escape pathway and to the continuum.

Understanding the formation of the quantum bridges seems paramount to optimising enhanced ionisation. Unfortunately, classical arguments are not sufficient in explaining the time evolution of the Wigner function during molecular enhanced ionisation. They fail to predict the cyclical motion of the quantum bridge after the bifurcation as well as its frequency. This all seems to point to that the evolution of the quantum bridge is inherently nonclassical. In order to quantify this we use the quantum Liouville equation, Eq.~(\ref{eq:liouville}) to define the amount of quantum corrections to the evolution of the Wigner function. The result is quite staggering and quantum corrections along with the corresponding Wigner quasiprobability distribution are shown in Fig.~\ref{fig:quant2}. If the Wigner function has a fully classical time evolution, $Q(x, p, t)$ vanishes everywhere. As shown in Fig.~\ref{fig:quant2} (b), the quantum corrections become very strong around the quantum bridge as it starts to build up. In contrast, they are completely absent along the semiclassical tail. For this we conclude that enhanced ionisation is due to the interplay of two pathways: The semiclassical escape pathways associated with tunneling mechanisms that follows the separatrix with a classical Wigner function evolution (despite its description of an inherently quantum mechanical process) and the quantum bridge. The latter stems from the interference between the two molecular wells, breaks all phase space constraints, presents very strong quantum corrections and cannot have a classical analogue.

\subsection{Understanding time scales}
\label{sub:timescales}

\begin{figure*}[tbp]
\centering
\includegraphics[width=12cm]{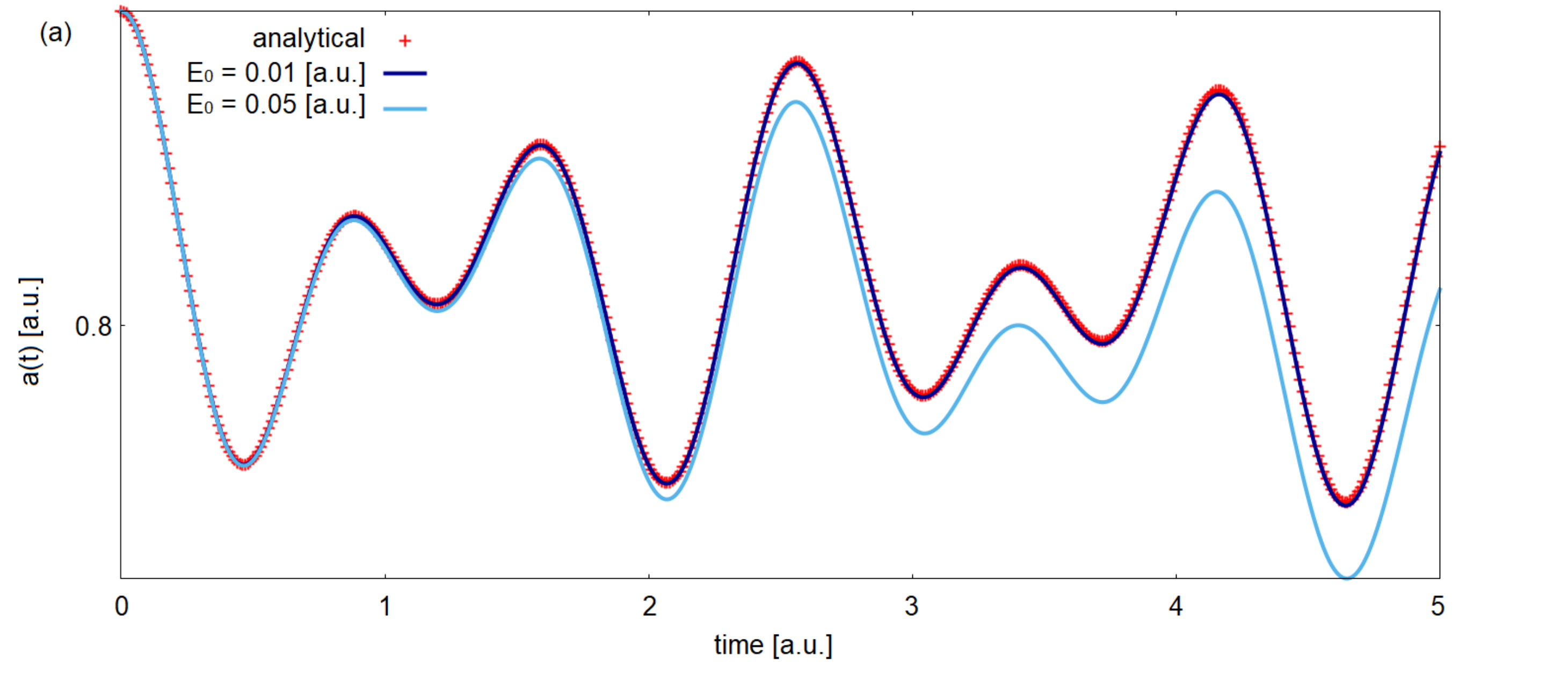}
\includegraphics[width=12cm]{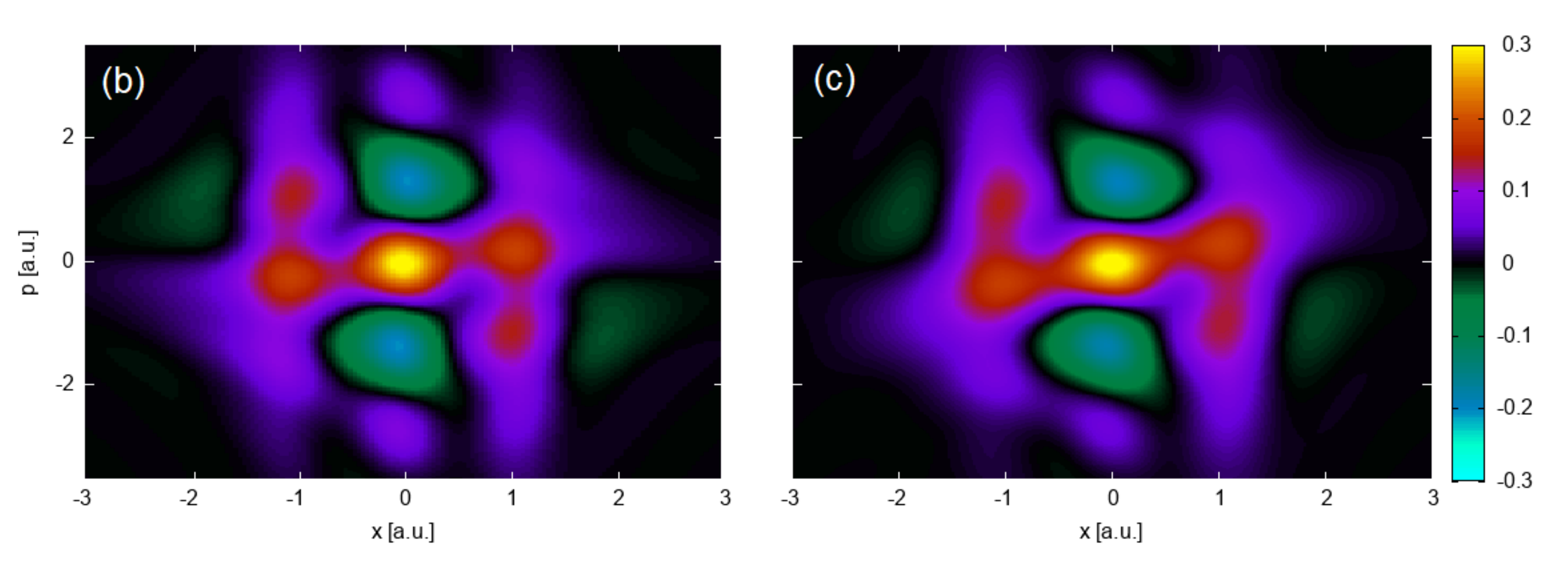}
\caption{(a) Comparison between the absolute value of the autocorrelation function $|a(t)|^2$, see Eq.~(\ref{eq:autocorrelation}), calculated using the analytical method in \cite{Kufel2020} in a field free system (red, dotted line) and numerical computations from \cite{Chomet2019} using the same parameters but with a static field of strength $\mathcal{E}_0 = 0.01$ a.u. ($I=3.51\times 10^{12} \mathrm{W}/\mathrm{cm}^2$) (dark blue solid line) and $\mathcal{E}_0 = 0.05$ a.u. ($I=1.72\times10^{14} \mathrm{W}/\mathrm{cm}^2$)(light blue solid line). [Bottom row] Wigner quasiprobability distributions using the same parameters as (a) at time $t = 0.7$ a.u. using in panel (b) the analytical method in \cite{Kufel2020} for a field free system and in panel (c) the numerical method in \cite{Chomet2019} for a static field of strength $\mathcal{E}_0 = 0.05$ a.u. ($I=1.7\times10^{14} \mathrm{W}/\mathrm{cm}^2$).}
\label{fig:analytical}
\end{figure*}

Phase space methods are also helpful for identifying time scales and the physics behind them. This is extremely important when working with time-dependent fields. For instance, in molecular enhanced ionisation, it is paramount to understand to timescale of formation of the quantum bridge as well as its cyclical motion, and how it relates to the frequency of the time-dependent field. Indeed, the quantum bridge can both aid or hinder ionisation and by changing parameters such as the type of field, the initial wave packet or the internuclear distance, enhanced ionisation can be optimised. 
As seen in \cite{Chomet2019}, the quantum bridge and its frequency is inherent to the molecular system, and is present even in the absence of an external field.  In \cite{Kufel2020} an analytical method based upon those employed in quasi-solvable models \cite{ushveridze1994,Fring2015,Fring2019} is used to determine the origin and the value of the frequency of the quantum bridge in a field-free system. Using the autocorrelation function of different initial wave packets, it is clear that this frequency is due to the coupling of different eigenstates. This leads to a very interesting situation in \cite{Chomet2019}, since the frequency of the quantum bridge is higher than the frequency of the time-dependent laser field. The quantum bridges could be controlled with the appropriate coherent superposition of states or different driving fields, which opens a wide range of possibilities for studying quantum effects in enhanced ionisation. We have verified that, for the model potential in \cite{Kufel2020}, the frequencies computed analytically are quite robust upon inclusion of an external static field. This is illustrated in Fig.~\ref{fig:analytical}\footnote{In these analytical calculations and their numerical counterparts, a parabolic potential model was employed instead of the soft-core and Gaussian potentials used in the remainder of the present work.}.

\begin{figure}[ht] 
\centering
    \includegraphics[width=11cm]{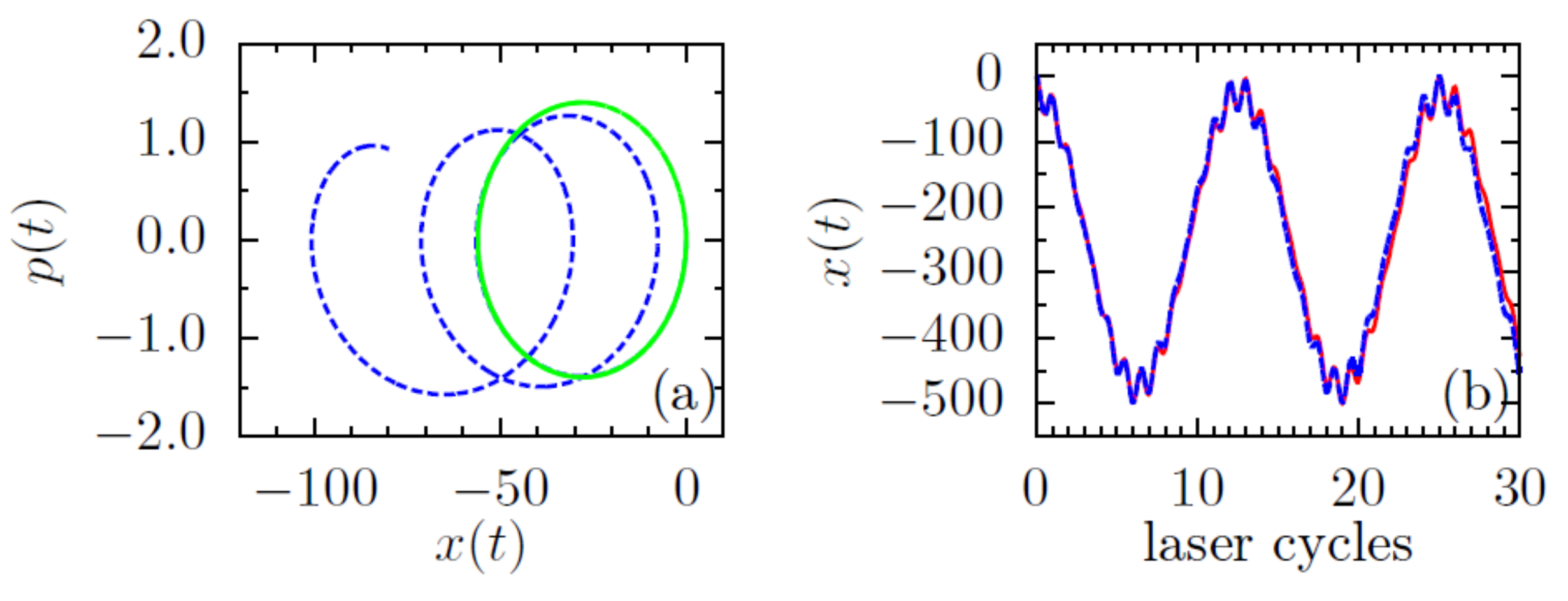}
    \caption{Numerical solutions of Mathieu's equation (dashed line) in phase space (panel (a)) and as a position-time plot for $\beta = 0.004$ (panel (b)). The initial positions and momenta in both cases is $x(0) = 0$ and $p(0) = 0$, respectively. For reference, the continuous line in panel (a) shows a closed orbit resulting from the propagation under a homogeneous laser field. The continuous line in panel (b) represents Dehmelt's approximation to Mathieu's equation. From \cite{Zagoya2016}.}
    \label{fig:time1}
\end{figure}

Another example are the several timescales that occur for HHG in inhomogeneous media \cite{Zagoya2016}. The very good agreement between the quantum time-frequency maps and the classical-trajectory computations means that we can explore the dynamical aspects of the latter. If the atomic potential $V_{sc}(x)$ is neglected, the equation of motion
\begin{equation}
	\label{eq:newtoneq}
	\frac{\mathrm{d}^2x}{\mathrm{d}t^2}=-\mathcal{E}(t)\beta\left(x+\frac{1}{\beta}\right)-\frac{\partial V_{sc}(x)}{\partial x},
\end{equation}
describing the trajectory ensemble in the inhomogeneous field can be re-written as Mathieu's equation. Explicitly,  
\begin{equation}
    \frac{d^2 \Upsilon}{d \tau ^2}+\epsilon \Upsilon cos \tau = 0,
\end{equation}
where  $\Upsilon= \beta x+1$, $\tau=\omega t$, $\epsilon = \beta \mathcal{E}_0 / \omega ^2$. This equation is widely used to study particles in ion traps \cite{Bluemel1989,Brkic2006,March1997,Paul1990}, and provides a wide range of dynamic information, from the values of the inhomogeneity parameter $\beta$ for which the system is stable, to the time scales involved. From the phase space study of individual trajectories, shown in Fig.~\ref{fig:time1}, one can see that they experience two differing motions: a slow and large oscillation and a small and rapid one. This allows us to apply Dehmelt's approximation within the stability region, and its accuracy is shown in Fig.~\ref{fig:time1}. From this we conclude that the secular oscillations are indeed responsible for the high-frequency structures in the time-resolved spectra seen in Fig.~\ref{fig:quant0}.

\section{Conclusions}
\label{sec:conclusions}

The take home message of the present review is that quantum optics, quantum information, chemical physics and the theory of dynamical systems have developed powerful toolkits that are under-used in strong-field and attosecond physics, among them classical and quantum phase space. We have provided a few examples of how phase space arguments and/or quantum quasiprobability distributions can be employed in the context of high-order harmonic generation and strong-field ionisation, be it for establishing constraints and determining different dynamical regimes, or for studying non-classical effects. Moreover, the phase space can also provide guidance for constructing effective Hamiltonians, determining relevant subspaces and understanding the interplay between the residual binding potentials and the laser field in greater depth. This is particularly important in the context of correlated multielectron dynamics, for which the large number of degrees of freedom may pose additional difficulties. Some of these techniques have been referred to in the traditional setting of laser-induced nonsequential double ionisation (see Sec.~\ref{sub:nsdi}), and will become increasingly  necessary for extended systems such as large molecules, solids and nanostructures. For instance, being able to select the relevant degrees of freedom and treating them quantum mechanically, while describing the less relevant ones classically, or incorporating quantum corrections around classical evolution are widespread strategies in quantum chemistry \cite{Miller2001,Miller2005}, cold gases \cite{Blakie2008,Polkovnikov2010} and in recent years photosynthetic compounds \cite{Tao2010,Teh2017}. 
Thereby, a key question is  how to adapt these techniques to a highly transient, subfemtosecond regime, in the context of attochemistry. Recently, quantum and classical approaches have been combined to investigate electron and nuclear dynamics in pump-probe experiments in glycine, and the initial state was computed using phase space techniques \cite{Delgado2021}.
Even in the single-electron regime, it is desirable to move from one-dimensional models. This interest ranges from a more accurate description of the dynamics \cite{Majorosi2018,Majorosi2020} to studies in orthogonally polarised fields \cite{Liu2021}. 

In addition to that, it is clear that one must go beyond traditional modelling in strong-field laser-matter interaction, which employs either pure quantum states or classical methods, and often takes the system to be initially in the ground state. One then assumes that the ionising electron creates a hole in the orbital with the lowest binding energy.  Nonetheless, one must bear in mind that, in real-life situations, the atomic and molecular ions generated by a strong laser field will be in a coherent superposition of states. This means that it is important to establish whether there will be coherences between the ionisation channels \cite{Rohringer2009}. Furthermore, the outgoing electron is expected to exhibit a degree of entanglement with its parent ion, which may harm coherence. This has been discussed in recent theoretical work, in which two time delayed XUV pulses are employed to control the degree of entanglement between vibrational and electronic degrees of freedom in $\mathrm{H}_2$. Entanglement prevents coherent  superpositions of states, which would lead to vibrational wave packets, from forming \cite{Vrakking2021}. To be able to determine and prepare the ion in appropriate coherent superposition is really important in the context of attosecond hole migration \cite{Goulielmakis2010,Pabst2011} and pump-probe schemes \cite{Sansone2010,Vrakking2021,Delgado2021}, which rely on well-defined phase relationships. Electronic coherences also play a key role in XUV induced bond formation \cite{Valentini2020}. Thereby, one must assess how nuclear and electronic degrees of freedom couple, with the aim of maximising coherence \cite{Goetz2016}. This also implies that a density-matrix formalism and effective Hamiltonians \cite{Rohringer2009,Law2018}, which are more suitable for open quantum systems, are required. 
This is particularly true if one takes into consideration the current trends, towards extended systems such as large molecules \cite{Remacle2006,Calegari2014,Lepine2014,Kuleff2016,Calegari2016Migration,Delgado2021} and nanostructures \cite{Ciappina2017,Ciappina2019}, for which overcoming decoherence, quantifying entanglement and non-classical behavior, and controlling the coupling with the environment will pose major challenges. 
There is theoretical evidence that nuclear degrees of freedom in a large molecule weaken coherence. Still, it is remarkable that even in the worst-case scenario phase relations may survive \cite{Delgado2021}. The present review is a brief illustration of how phase space tools widely used in other research areas, such as quasiprobability densities, may be employed in attosecond physics. 

We would like to thank C. Hofmann and A. Palacios for useful discussions, A. Serafini and S. Bose for providing references and acknowledge funding from the UK Engineering and Physical Sciences Research Council (EPSRC), grant no. EP/J019143/1.

\printbibliography

\end{document}